\documentclass[bimj,fleqn]{w-art}
\usepackage{times}
\usepackage{w-thm}
\usepackage[authoryear]{natbib}
\usepackage{booktabs}
\usepackage[colorlinks,bookmarksopen,bookmarksnumbered,citecolor=black,urlcolor=black]{hyperref}

\newcommand{\bcx}{{\bf X}}
\newcommand{\bcy}{{\bf Y}}

\newcommand{\bcv}{{\bf V}}
\newcommand{\bcw}{{\bf W}}
\newcommand{\bci}{{\bf I}}
\newcommand{\bch}{{\bf H}}

\newcommand{\bcm}{{\bf M}}

\newcommand{\bcd}{{\bf D}}

\newcommand{\br}{{\bf r}}

\newcommand{\bmu}{\mbox{\boldmath $\mu$}}

\newcommand{\bzero}{{\bf 0}}

\newcommand{\bomega}{\mbox{\boldmath $\Omega$}}

\theoremstyle{plain}

\theoremstyle{definition}

\usepackage{enumitem}
\usepackage[]{graphicx}
\chardef\bslash=`\\ % p. 424, TeXbook

\begin{document}
%\DOIsuffix{bimj.DOIsuffix}

%\DOIsuffix{bimj.200100000}
%\Volume{52}
%\Issue{61}
%\Year{2010}
%\pagespan{1}{}
\keywords{Covariate adjustment; Finite-sample corrections; Generalized estimating equations; Inverse probability weights; Overlap weights;  Sandwich variance estimator;\\
%\noindent \hspace*{-4pc} {\small\it (Up to five keywords are allowed and should be given in alphabetical order. Please capitalize the key}\\
%\hspace*{-4pc} {\small\it words)}\\[1pc]
%\noindent\hspace*{-4.2pc} Supporting Information for this article is available from the author or on the WWW under\break \hspace*{-4pc} \underline{http://dx.doi.org/10.1022/bimj.XXXXXXX} (please delete if not applicable)
} %%% semicolon and fullpoint added here for keyword style

% \title[Individual-level covariate adjustment in small CRTs with a rare outcome]%{Leveraging baseline covariates to analyze small cluster-randomized trials with a rare binary outcome}
% {Leveraging \textcolor{black}{individual-level} covariates to analyze small cluster-randomized trials with a rare binary outcome}

\title[Covariate adjustment in small CRTs with a rare outcome]{Leveraging baseline covariates to analyze small cluster-randomized trials with a rare binary outcome}

%% Information for the first author.
\author[Zhu {\it{et al.}}]{Angela Y. Zhu\footnote{Corresponding author: {\sf{e-mail: azhu147@its.jnj.com}}}\inst{,1}} 
\address[\inst{1}]{{Department of Biostatistics, Epidemiology, and Informatics}, {University of Pennsylvania Perelman School of Medicine}, {{Philadelphia, PA 19104}, {United States of America}}}
\author[]{Nandita Mitra\inst{1}}
\author[]{Karla Hemming\inst{2}}
\address[\inst{2}]{{Department of Public Health, Epidemiology, and Biostatistics}, {University of Birmingham Institute of Applied Health Research}, {{Birmingham B15 2TT}, {United Kingdom}}}
\author[]{Michael O. Harhay\footnote{Co-senior authors}\inst{,1}}
\author[]{Fan Li\inst{**,3,4}}
\address[\inst{3}]{{Department of Biostatistics}, {Yale School of Public Health}, {{New Haven, CT 06510}, {United States of America}}}
\address[\inst{4}]{{Center for Methods in Implementation and Prevention Science}, {Yale School of Public Health}, {{New Haven, CT 06510}, {United States of America}}}

%\Receiveddate{zzz} \Reviseddate{zzz} \Accepteddate{zzz} 

\begin{abstract}
Cluster-randomized trials (CRTs) involve randomizing entire groups of participants---called clusters---to treatment arms but are often comprised of a limited or fixed number of available clusters. While covariate adjustment can account for chance imbalances between treatment arms and increase statistical efficiency in individually-randomized trials, analytical methods for \textcolor{black}{individual-level} covariate adjustment in small CRTs have received little attention to date. In this paper, we systematically investigate, through extensive simulations, the operating characteristics of propensity score weighting and multivariable regression as two \textcolor{black}{individual-level} covariate adjustment strategies for estimating the participant-average causal effect in small CRTs with a rare binary outcome and identify scenarios where each adjustment strategy has a relative efficiency advantage over the other to make practical recommendations. We also examine the finite-sample performance of the bias-corrected sandwich variance estimators associated with propensity score weighting and multivariable regression for quantifying the uncertainty in estimating the participant-average treatment effect. To illustrate the methods for individual-level covariate adjustment, we reanalyze a recent CRT testing a sedation protocol in $31$ pediatric intensive care units. 
\end{abstract}

%% maketitle must follow the abstract.
\maketitle              % Produces the title.

\section{Introduction}

In cluster-randomized trials (CRTs), \textcolor{black}{entire clusters} of subjects, rather than the individuals themselves, are randomly allocated to treatment and control arms \citep{turner2017a}. Examples of clusters include hospitals, schools, and residential care homes \citep{Murray1998}. CRTs are essential when an intervention is delivered at the cluster level, and are often also used when there are concerns over contamination or because of their logistical appeal. \textcolor{black}{Due to cluster randomization and the fact that individuals within each cluster share the same physical or social factors, the intracluster correlation coefficient (ICC) that describes the similarity between outcomes within the same cluster plays a central role in the design and analysis of CRTs.} This ICC must be accounted for in analysis to avoid inflated type I error rates \citep{turner2017b}.

Methods to analyze CRTs include cluster-level analyses that use summary measures for each cluster and individual-level analyses that employ generalized linear mixed models or marginal models estimated with  generalized estimating equations (GEEs) \citep{Liang1986}. Marginal models are sometimes preferred because the interpretation of the intervention effect parameter does not depend on the assumed correlation structure \citep{Preisser2003}. In addition, marginal models coupled with the sandwich variance estimator have been shown to be robust to misspecification of the working covariance structure and to provide asymptotically valid confidence intervals as long as the marginal mean structure is correctly specified \citep{Liang1986}. However, when the number of clusters is small (typically not exceeding 30), the sandwich variance estimator can exhibit negative bias and may lead to under-coverage of the associated confidence interval estimator. Indeed, \citet{ivers2011} reviewed 300 CRTs published between 2000 and 2008 and found the median number of clusters to be 21. Given small CRTs are common, there remains considerable interest in improving finite-sample inference of the sandwich variance estimator for GEE analysis such trials \citep{leyrat2018}. 

Previous simulation studies have compared the performances of several bias-corrected sandwich variance estimators under certain settings. Specifically, \citet{lu2007} compared the bias-corrected sandwich variance estimators due to \citet{kauermann2001} (KC) and \citet{mancl2001} (MD). They found that the normal confidence interval estimator with the MD variance formula often leads to coverage of 95\% confidence intervals near the nominal level. \citet{liRedden2015} have recommended the corrections proposed by KC and \citet{fay2001} (FG) depending on variations of cluster size when analyzing binary outcomes. \citet{ford2017} have indicated that the KC bias-corrected sandwich variance estimator may still give downwardly biased estimates of the standard error and that the FG bias-corrected sandwich variance estimator performs similarly to that of KC. Further, they note that the MD bias-corrected sandwich variance estimator tends to over-correct resulting in conservative inference, and thus recommended the average of the MD and KC standard error estimators as the top performer in CRTs with continuous and binary outcomes. Similar observations and recommendations were discussed in \citet{li2021sample} for CRTs with count outcomes subject to truncation. It is evident that the performance of these bias-corrected sandwich variance estimators can depend on the settings of interest. 

\textcolor{black}{One limitation of existing comparative studies of analytical methods for CRTs is that they have primarily focused on unadjusted analyses without covariates or adjusted analyses with cluster-level covariates. For example, \citet{westgate2013small} and \citet{ford2017} have investigated the small-sample corrections in the presence of cluster-level covariates. They pointed out that the performance of MD and KC bias-corrected sandwich variance estimators may deteriorate with cluster-level covariates, while the average of MD and KC standard error estimator was relatively more robust. However, the performance of small-sample corrections remains to be explored in settings where adjustment for individual-level covariates is being considered.} In CRTs, \textcolor{black}{individual-level covariates} are frequently collected at baseline, and the need for covariate adjustment can fall into one of the following categories. First, in CRTs where cluster-level aggregates of \textcolor{black}{individual-level covariates} are utilized during restricted randomization, adjusting for such \textcolor{black}{individual-level covariates} in the analysis model can adequately control for the type I error rate \citep{Li2015,Li2017,watson2021design,zhou2021constrained}. Second, adjusting for \textcolor{black}{individual-level covariates} can be based on precision considerations, analogous to the justifications provided in individually-randomized trials \citep{williamson2014variance,zeng2021propensity}. Third, \textcolor{black}{individual-level covariate} adjustment may reduce selection or recruitment bias compared to the unadjusted analysis in CRTs \citep{leyrat2013propensity,leyrat2014propensity,li2021clarifying}. 

\textcolor{black}{In this article, we design a neutral simulation study to compare methods for analyzing small CRTs in the presence of individual-level covariate adjustment. We focus on a rare binary outcome (event rate $\leq 10\%$ as in our motivating study in Section \ref{sec:Data}). Besides considering individual-level covariate adjustment, the unique innovations of our simulation study are three-fold. First, we recognize the potential ambiguity in the treatment effect parameter under individual-level covariate adjustment for binary outcomes (due to non-collapsibility) and so employ the potential outcomes framework to define a participant-average treatment effect on the odds ratio scale as a common target estimand. Fixing a common target estimand has been recommended for individually-randomized trials \citep{benkeser2021} and can facilitate an objective comparison across different methods where the associated regression coefficients may have different interpretations. This work thus represents a first effort to operationalize this important consideration for CRTs. Second, we focus on a rare outcome where adjusting for multiple individual-level covariates may lead to separation issues and non-convergence when fitting multivariable regression models. Complete separation occurs when the value of the outcome may be determined or distinguished for values of the predictors; that is, there is no observed variability in the outcome for certain combinations of covariate values. This convergence problem, however, may be ameliorated using propensity score weighting \citep{williamson2014variance,zeng2021propensity}, which has been demonstrated to be a valid covariate adjustment approach in individually-randomized trials, and we adapt this approach to estimate causal effects in CRTs. Finally, under a GEE modeling framework, we contribute new evidence on whether the bias-corrected sandwich variance estimators can provide valid inference for the participant-average treatment effect under individual-level covariate adjustment when the number of clusters is small.}

\section{Methods for \textcolor{black}{Individual-Level} Covariate Adjustment in CRTs with a Rare Binary Outcome} \label{sec:Methods}

\subsection{Notation and estimand}\label{sec:estimand}
Suppose we have a two-arm parallel CRT with $N$ total clusters and 1:1 randomization, and let $Z$ denote the treatment indicator with $Z=1$ corresponding to inclusion in the treatment group and $Z=0$ corresponding to being in the control group. Let $Y_{ij}$ denote the binary outcome of the $j$th subject in cluster $i$ with $P$-dimensional covariate vector $\bcx_{ij}$, $i=1,\ldots,N$ and $j=1,\ldots,m_i$. We assume the outcomes are correlated within the same cluster but independent across clusters. Specifically, for the $i$th cluster, $\bcy_i = [Y_{i1},\ldots,Y_{im_i}]'$ is the outcome vector and $\bcx_i = [\bcx_{i1},\ldots,\bcx_{im_i}]'$ is the covariate matrix. In this article, we focus on estimating an average treatment effect on the ratio scale. Briefly, as the outcome of interest is binary, we describe a target estimand as the log causal odds ratio of the treatment to the control. We \textcolor{black}{proceed under the potential outcomes framework and} assume $\{Y_{ij}(1),Y_{ij}(0)\}\in\{0,1\}^{\otimes 2}$ as the pair of potential/counterfactual binary outcomes under the treatment and control conditions; for example, $Y_{ij}(1)$ is the binary potential outcome of the $j$th subject in cluster $i$ had cluster $i$ been randomized to treatment, and $Y_{ij}(0)$ is the binary potential outcome of the $j$th subject in cluster $i$ had cluster $i$ been randomized to control. \textcolor{black}{We consider a sampling-based framework where the observed clusters are assumed to be a random sample of a population of clusters with randomly varying cluster sizes. This framework allows us to define the expectation sign over the distribution of clusters and permits us to formalize the target estimand---participant-average treatment effect on the odds ratio scale---as}
\begin{equation}\label{eq:OR}
    OR = \frac{E\left(\sum_{j=1}^{m_i} Y_{ij}(1)\right) E\left(m_i-\sum_{j=1}^{m_i} Y_{ij}(0)\right)}{E\left(\sum_{j=1}^{m_i} Y_{ij}(0)\right)E\left(m_i-\sum_{j=1}^{m_i} Y_{ij}(1)\right)},
\end{equation}
and we further define $\Delta=\log(OR)$ as our target parameter in the ensuing simulations. While our focus is on estimand \eqref{eq:OR}, we refer to \citet{Brennan2022IJE}, \citet{wang2022two}, and \citet{su2021model} for alternative estimands such as the cluster-average treatment effect or other weighted cluster-average treatment effects that may also be of interest in CRTs. 

\subsection{Overview of generalized estimating equations (GEE) analyses of CRTs}

To estimate the parameter defined in \eqref{eq:OR}, we primarily consider the GEE approach, under which, the relationship between the marginal mean $E[Y_{ij}|\bcx_{ij}] = \mu_{ij}$ and the covariates $\bcx_{ij}$ may be represented by a generalized linear model, $g(\mu_{ij}) = \bcx_{ij}\boldsymbol{\beta}$. With a binary outcome, logistic regression is often used where $g$ is taken as the canonical logit link function, which we use in this article. Let $\bcv_i$ denote the working covariance structure for $\bcy_i$. The parameter estimator $\hat{\boldsymbol{\beta}}$ in the marginal model is consistent and asymptotically normal, and, even if the working correlation structure is misspecified, its variance-covariance can be consistently estimated by
$\bcv = Cov(\hat{\boldsymbol{\beta}}) = \hat{\bomega} \left(\sum_{i=1}^N \bcd_i'\bcv_i^{-1}\br_i\br_i'\bcv_i^{-1}\bcd_i \right)\hat{\bomega}$, where $\bcd_i={\partial \bmu_i}/{\partial \boldsymbol{\beta}'}$, $\hat{\bomega} = \left(\sum_{i=1}^N \bcd_i'\bcv_i^{-1}\bcd_i \right)^{-1}$ and $\br_i\br_i' = (\bcy_i-\hat{\bmu}_i)(\bcy_i-\hat{\bmu}_i)'$ is an estimate of the covariance of $\bcy_i$. The variance $\bcv$ is often referred to as the robust sandwich estimator or the empirical sandwich estimator \citep{Liang1986}. 

Frequently, the primary analysis of CRTs proceeds with the marginal model without any baseline covariates---the so-called unadjusted analysis. In this approach, $\bcx_{ij}=Z_i$ in the marginal mean model and the mean model can be explicitly written as 
\begin{equation}\label{eq:unadj}
\text{logit}(\mu_{ij})=\beta_0+\beta_Z Z_i,
\end{equation}
and $\boldsymbol{\beta}=(\beta_0,\beta_Z)'$. To estimate $\boldsymbol{\beta}$, one may choose either the independence or exchangeable working correlation structure. \textcolor{black}{When the true outcome data generating model is indeed the unadjusted model without any further covariates \eqref{eq:unadj}, $\beta_Z$ directly corresponds to our target estimand $\Delta$, and the choice of exchangeable working correlation structure corresponds to modeling the ICC and can often provide a more efficient average causal effect estimator when the cluster sizes are variable \citep{li2021sample,li2021sample_bin}. In this case, even when the exchangeable working structure is not the true correlation structure, GEE models are robust to such misspecification.} However, when the true data generating process involves additional individual-level covariates or when the cluster size is predictive of the treatment effect, we show in the Supplementary File Section 1 that $\hat{\beta}_Z$ is a consistent estimator to $\Delta$ under the independence working correlation structure but often not otherwise. This argument extends the ones provided in \citet{wang2022two} and \citet{Brennan2022IJE} to ratio effect measures. Because of this rationale, we primarily focus on the case with an independence working correlation model. Beyond the unadjusted analysis, the GEE approach can be extended to leverage baseline individual-level covariates to potentially increase the efficiency for estimating $\Delta$. In Section \ref{sec:PS}, we consider using propensity score weighting for covariate adjustment. In addition, model \eqref{eq:unadj} can be expanded to include additional baseline covariates, in which case a population standardization procedure is required to estimate $\Delta$, because the regression coefficient does not directly correspond to $\Delta$ due to non-collapsibility. We describe this multivariable regression approach in Section \ref{sec:multivar}. \textcolor{black}{As CRTs often include a limited number of clusters (typically not exceeding 30), a set of bias-corrected sandwich variance estimators to address the finite-sample bias in inference is also described below.}

\subsection{Propensity score weighting for \textcolor{black}{individual-level} covariate adjustment}\label{sec:PS}
\citet{rosenbaum1983central} developed the propensity score, which is defined as the probability of treatment conditional on observed covariates; that is,  $e(\bcx) = P(Z=1|\bcx)$. Approaches based on the propensity score, such as matching, weighting, and stratification, are commonly employed in the design and analysis of observational studies to control for confounding, since it has been shown that conditional on the propensity score, treatment is randomized. In randomized trials, the true propensity score is known by design and there is no need to model the propensity score for unbiased estimation of treatment effect. However, propensity score weighting has been shown to provide efficiency gains by addressing chance imbalance of baseline covariates. \textcolor{black}{\citet{williamson2014variance} have shown in individually-randomized trials that (i) inversely weighting by the estimated propensity score with prognostic covariates can reduce the variance of the unadjusted treatment effect estimator without compromising the population-level causal estimand, and (ii) with a rare binary outcome, the propensity score weighting approach often circumvents non-convergence issues that multivariable regression is vulnerable to.} In addition, \citet{zeng2021propensity} demonstrated that in individually-randomized trials, weighting by overlap weights almost always leads to smaller variance than inverse propensity score weighting. \textcolor{black}{Compared to multivariable regression adjustment for covariates, propensity score weighting has the additional benefit of preserving the population-level causal estimands but have not yet been explored in CRTs as a potentially effective tool for individual-level covariate adjustment. Here, we explore the use of these two propensity score weighting approaches as individual-level covariate adjustment strategies for CRTs, without compromising the clarity of the target estimand.} 

Inverse probability of treatment weighting (IPW) seeks to construct a sample in which the distributions of observed baseline variables are similar between treatment and control groups. These weights are defined to be the reciprocal of the conditional probability of being assigned to the treatment group that they were observed to be in. In CRTs, for subject $j$ in cluster $i$, the weight is given by,
\begin{equation*}
    w_{ij} = 
    \begin{cases}
    {1}/{e(\bcx_{ij})} & \text{ if treated } (Z_{i}=1) \\
    {1}/\{1-e(\bcx_{ij})\} & \text{ if control } (Z_{i}=0)
    \end{cases}
\end{equation*}
On the other hand, overlap weighting (OW) was proposed to overcome possible limitations of IPW when there is limited overlap in covariate distributions between treatment arms in observational studies \citep{li2018} and to also improve upon IPW in individually-randomized trials. Specifically, the overlap weights are defined to be the probability of being in the opposite treatment group (the one the subject was not observed to be in) given baseline covariates:
\begin{equation*}
    w_{ij} = 
    \begin{cases}
    1-e(\bcx_{ij}) & \text{ if treated } (Z_{i}=1) \\
    e(\bcx_{ij}) & \text{ if control } (Z_{i}=0)
    \end{cases}
\end{equation*}
\textcolor{black}{For individual-level randomized trials, the true propensity score is usually a constant and does not depend on individual-level covariates; therefore, IPW and OW correspond to the same population estimand \citep{zeng2021propensity}, but previous simulations indicate that OW provides better finite-sample performance in that it is more efficient at smaller sample sizes. Under cluster randomization, the true propensity score is determined by study design and does not depend on individual-level covariates. Therefore, using OW and IPW can both target the participant-average treatment effect \eqref{eq:OR} as a common estimand.}
%but there may be differences in finite-sample efficiency, which we will investigate.}   

The propensity score is often estimated using parametric logistic regression.  Alternative models for propensity score estimation that have been considered include neural nets, decision trees, Bayesian additive regression trees (BART), and ensemble learners \citep{westreich2010, zhu2021}. In observational studies, these nonparametric machine learning approaches can provide greater flexibility and more accurate estimates when there are complex relationships among the variables; however, their role in the analysis of CRTs remains under-explored. \textcolor{black}{In particular, BART is a Bayesian nonparametric sum-of-trees model that involves binary splits in the predictor space and a regularization prior to avoid overfitting \citep{chipman2010}. This model has become popular for estimating treatment effects in observational studies due to its computational efficiency and flexibility in modeling complex relationships between variables \citep{hill2011}. In settings where there are nonlinear relationships or interactions among the covariates in the propensity score or outcome models, there may be difficulty in specifying those patterns through a simple parametric model. BART models can handle the inclusion of several predictors, allow for nonlinear relationships and interactions between predictors, and have been demonstrated to be effective in past simulations for observational studies \citep{dorie2019automated}.} In the ensuing simulations, we explore whether using BART to estimate propensity scores will impact results in the CRT context.

Next, we describe the point estimator and the bias-corrected sandwich variance estimators under propensity score weighted GEE. The propensity score weighted GEE is essentially based on the same marginal mean model as in \eqref{eq:unadj}, but including propensity score weights in the estimating equations for $\boldsymbol{\beta}$. Specifically, once the propensity scores are estimated for each subject, the weight matrix $\bcw_i$ is formed for each cluster $i$ with $w_{ij}$ on the diagonal and 0 for the off diagonal elements. The regression parameters $\boldsymbol{\beta}=(\beta_0,\beta_Z)'$ are estimated by solving the weighted GEE, $\sum_{i=1}^N \bcd_i'\bcv_i^{-1}\bcw_i (\bcy_i-\bmu_i) = \bzero$, where $\bmu_i=(\mu_{i1},\ldots,\mu_{im_i})'$,  
$$\mu_{ij}=\frac{\exp(\beta_0+\beta_Z Z_i)}{1+\exp(\beta_0+\beta_Z Z_i)},$$
and $\bcv_i$ is the working variance under the independence working correlation model.

Under the independence working correlation assumption, since the weighted GEE model only contains the treatment indicator as a covariate and the weights are only used to control for chance imbalance, the treatment coefficient estimator is the $\hat{\beta}_Z=log(\widehat{OR})$, which is a consistent estimator for the participant-average treatment effect $\Delta$. For convenience, we ignore the variability due to the estimation of propensity scores, and the corresponding bias-corrected sandwich variance estimators incorporating these weights are developed in Table \ref{tab:estimators}(a). Here, $\hat{\bomega} = (\sum_{i=1}^N \bcd_i' \bcv_i^{-1} \bcw_i \bcd_i)^{-1}$ is the propensity score weighted ``model-based'' variance, and $\bch_i= \bcd_i\hat{\bomega}\bcd_i'\bcv_i^{-1}  \bcw_i$ is the propensity score weighted leverage matrix for cluster $i$.
%Further, for the FG bias-corrected sandwich variance estimator, $\bcf_i=diag\{(1-min\{.75, [\bcq_i]_{jj}\})^{-1/2}\}$ and $\bcq_i = \bcd_i' \bcv_i^{-1} \bcw_i \bcd_i\hat{\bomega}$, which also includes the weight matrix with estimated propensity scores. 
%Note that the sandwich variance estimators here differ from previous bias-correction forms in that weight matrices have been integrated. 
\textcolor{black}{Although we focus on the MD and KC bias-corrected variance estimators in our simulations, prior studies have indicated that the FG bias-corrected variance estimator had very similar performance to the KC bias-corrected variance estimator \citep{li2018sample}. In fact, without weighting, \citet{scott2017finite} and \citet{wang2022power} have shown that the KC bias-corrected variance is equivalent to a modified version of the FG bias-corrected variance. This analytical insight also holds in the presence of weighting. We therefore do not further consider the FG bias-correction due to its similarity to the KC bias-correction, but still provide code to implement the FG bias-correction in the Supplementary Material.}
%and we will examine whether any of these bias-corrections can help maintain the nominal coverage of the covariate-adjusted estimation of $\Delta$ when there is only a limited number of clusters. 

\begin{table}[htbp]
\caption{Bias-corrected sandwich variance estimators for propensity score weighted GEE and multivariable adjusted GEE estimators of the participant average treatment effect. Notice that under propensity score weighting, $Cov (\hat{\boldsymbol{\beta}})$ is a $2\times 2$ matrix and the variance estimator for $\hat{\beta}_Z=log(\widehat{OR})$ is the lower-right element of $Cov (\hat{\boldsymbol{\beta}})$. \textcolor{black}{The matrix $\bcm$ is defined in equation \eqref{eq:M}.}}
\label{tab:estimators}
\begin{tabular}{lc}
\toprule
\multicolumn{2}{c}{(a) Variance estimation under propensity score weighting}                                                                                                                                                                                                                            \\ \toprule
\multicolumn{1}{c}{\textbf{Estimator}}                                                                                     & $Cov (\hat{\boldsymbol{\beta}})$                                                                                                                                      \\ \midrule
\multicolumn{1}{c}{Robust sandwich estimator}                                                                     & $\hat{\bomega}\left\{\sum_{i=1}^N\bcd_i' \bcv_i^{-1} \bcw_i \br_i \br_i' \bcw_i \bcv_i^{-1} \bcd_i\right\}\hat{\bomega}$                                              \\ \midrule
\multicolumn{1}{c}{\begin{tabular}[c]{@{}c@{}}Mancl and DeRouen\\ (MD) bias-corrected estimator\end{tabular}}     & $\hat{\bomega} \left\{\sum_{i=1}^N \bcd_i' \bcv_i^{-1} \bcw_i (\bci-\bch_i)^{-1} \br_i \br_i' (\bci-\bch_i')^{-1} \bcw_i \bcv_i^{-1}\bcd_i\right\}\hat{\bomega}$      \\ \hline
\multicolumn{1}{c}{\begin{tabular}[c]{@{}c@{}}Kauermann and Carroll\\ (KC) bias-corrected estimator\end{tabular}} & $\hat{\bomega} \left\{\sum_{i=1}^N \bcd_i' \bcv_i^{-1} \bcw_i (\bci-\bch_i)^{-1/2} \br_i \br_i' (\bci-\bch_i')^{-1/2} \bcw_i \bcv_i^{-1} \bcd_i\right\}\hat{\bomega}$ \\ 
%\midrule
%\multicolumn{1}{c}{\begin{tabular}[c]{@{}c@{}}Fay and Graubard\\ (FG) bias-corrected estimator\end{tabular}}      & $\hat{\bomega} \left\{\sum_{i=1}^N \bcf_i \bcd_i' \bcv_i^{-1} \bcw_i \br_i \br_i' \bcw_i \bcv_i^{-1} \bcd_i \bcf_i\right\}\hat{\bomega}$                              \\ %\bottomrule
\toprule
\multicolumn{2}{c}{(b) Variance estimation under a multivariable model}                                                                                                                                                                                                                                  \\ \midrule
\multicolumn{1}{c}{\textbf{Estimator}}                                                                                     & $\widehat{Var} (log (\widehat{OR}))$                                                                                                                                  \\ \midrule
\multicolumn{1}{c}{Robust sandwich estimator}                                                                     & $\bcm' \hat{\bomega}\left\{\sum_{i=1}^N \bcd_i' \bcv_i^{-1} \br_i \br_i'\bcv_i^{-1} \bcd_i\right\}\hat{\bomega}\bcm$                                                  \\ \midrule
\multicolumn{1}{c}{\begin{tabular}[c]{@{}c@{}}Mancl and DeRouen\\ (MD) bias-corrected estimator\end{tabular}}     & $\bcm' \hat{\bomega} \left\{\sum_{i=1}^N \bcd_i' \bcv_i^{-1} (\bci-\bch_i)^{-1} \br_i \br_i' (\bci-\bch_i')^{-1} \bcv_i^{-1}\bcd_i\right\}\hat{\bomega} \bcm$         \\ \midrule
\multicolumn{1}{c}{\begin{tabular}[c]{@{}c@{}}Kauermann and Carroll\\ (KC) bias-corrected estimator\end{tabular}} & $\bcm' \hat{\bomega} \left\{\sum_{i=1}^N \bcd_i' \bcv_i^{-1} (\bci-\bch_i)^{-1/2} \br_i \br_i' (\bci-\bch_i')^{-1/2} \bcv_i^{-1} \bcd_i\right\}\hat{\bomega} \bcm$    \\ 
%\midrule
%\multicolumn{1}{c}{\begin{tabular}[c]{@{}c@{}}Fay and Graubard\\ (FG) bias-corrected estimator\end{tabular}}      & $\bcm' \hat{\bomega} \left\{\sum_{i=1}^N \bcf_i \bcd_i' \bcv_i^{-1} \br_i \br_i' \bcv_i^{-1} \bcd_i \bcf_i\right\}\hat{\bomega} \bcm$                                 \\ 
\bottomrule
\end{tabular}
\end{table}

\subsection{Direct multivariable regression for \textcolor{black}{individual-level} covariate adjustment}\label{sec:multivar}

As an alternative to propensity score weighting, we also consider the direct multivariable regression approach for covariate adjustment in CRTs. Due to non-collapsibility with the logit link function, the estimator of the coefficient for the treatment variable (when there are other individual-level covariates in the model) only reflects a conditional treatment effect. To ensure we are targeting the correct estimand $\Delta$, we will use the model fit to obtain estimates of the probability of the potential outcomes averaged or standardized over the covariate distribution. Suppose we fit the multivariable model $logit \left\{E(Y_{ij}|X_{ij}, Z_i)\right\} = \beta_0 + \sum_{p=1}^P \beta_p X_{ij}^{(p)} + \beta_Z Z_i $. Let $\hat{\boldsymbol{\beta}}' = [\hat{\beta}_0, \hat{\beta}_1,...,\hat{\beta}_P, \hat{\beta}_Z]$ denote the estimates of the coefficients from the GEE model fit, where $\hat{\delta}$ represents the log odds ratio estimator conditional on all covariates. \textcolor{black}{For estimating the multivariable adjusted GEE model, one could choose the working correlation to be either the independence correlation or the exchangeable correlation, and we focus \textcolor{black}{on} the former in this paper for an objective comparison with the propensity score approaches.} Based on the GEE model fit, the predicted risk for subject $j$ in cluster $i$, if he/she is given treatment is $\hat{P}_{1, ij} = {exp(\bcx_{1, ij}'\hat{\boldsymbol\beta})}/\left\{1+exp(\bcx_{1,ij}'\hat{\boldsymbol\beta})\right\}$, where $\bcx_{1, ij} = [1, \bcx_{ij}', 1]'$. Similarly, the predicted risk for subject $j$ in cluster $i$ under the control is $\hat{P}_{0, ij} = {exp(\bcx_{0, ij}'\hat{\boldsymbol\beta})}/\left\{1+exp(\bcx_{0, ij}'\hat{\boldsymbol\beta})\right\}$
where $\bcx_{0, ij} = [1, \bcx_{ij}', 0]'$. Then a standardization estimator (or referred to as the g-computation formula estimator) of the participant-average treatment effect is
\begin{equation*}
    log (\widehat{OR}) = log \left\{ \frac{\left(\sum_{i=1}^N\sum_{j=1}^{m_i} \hat{P}_{1,ij}\right) \left(\sum_{i=1}^N\left(m_i-\sum_{j=1}^{m_i} \hat{P}_{0,ij}\right)\right)}{\left(\sum_{i=1}^N\sum_{j=1}^{m_i} \hat{P}_{0,ij}\right)\left(\sum_{i=1}^N\left(m_i-\sum_{j=1}^{m_i} \hat{P}_{1,ij}\right)\right)}\right\}
\end{equation*}
For variance estimation, using the delta method, we arrive at $\widehat{Var} (log (\widehat{OR}))= \bcm' Cov(\hat{\boldsymbol{\beta}}) \bcm$, where $Cov(\hat{\boldsymbol{\beta}})$ may either be the robust sandwich estimator or the bias-corrected estimators in the existing literature \citep{Li2015,ford2017} and $\bcm = \partial log (\widehat{OR})/\partial\boldsymbol{\beta}$. We provide detailed steps of the derivation in Supplementary File Section 2 to show that $\bcm$ takes the form of 
\begin{align}\label{eq:M}
    \bcm =& \left(\frac{1}{\sum_{i=1}^N\sum_{j=1}^{m_i} \hat{P}_{1,ij}} + \frac{1}{\sum_{i=1}^N\left(m_i-\sum_{j=1}^{m_i} \hat{P}_{1,ij}\right)} \right) \sum_{i=1}^N\sum_{j=1}^{m_i} \frac{exp(\bcx_{1,ij}'\hat{\boldsymbol{\beta}})\bcx_{1,ij}}{\{1+exp(\bcx_{1,ij}'\hat{\boldsymbol{\beta}})\}^2}- \nonumber \\
    & \left(\frac{1}{\sum_{i=1}^N\sum_{j=1}^{m_i} \hat{P}_{0,ij}} + \frac{1}{\sum_{i=1}^N\left(m_i-\sum_{j=1}^{m_i} \hat{P}_{0,ij}\right)} \right) \sum_{i=1}^N\sum_{j=1}^{m_i} \frac{exp(\bcx_{0,ij}'\hat{\boldsymbol{\beta}})\bcx_{0,ij}}{\{1+exp(\bcx_{0,ij}'\hat{\boldsymbol{\beta}})\}^2} 
\end{align}
The analogous robust sandwich estimator and bias-corrected sandwich estimators are also summarized in Table \ref{tab:estimators}(b) for quick reference.

\section{Simulation Studies} \label{sec:Simulations}
%We use simulation studies to evaluate the performance and properties of GEE estimators described above \textcolor{black}{for estimating the same target estimand} under a wide range of scenarios. 
We employ the ADEMP (aims, data-generating mechanisms, estimands, methods, and performance measures) structured approach proposed by \citet{morris2019} to describe the details of our simulation studies.

\subsection{Aims}
The motivation of our simulation studies is to inform practical choices for individual-level covariate-adjusted analysis of CRTs with small numbers of clusters and rare binary outcomes. 
%These studies are intended to provide evidence-supported guidance in potentially challenging scenarios for the methods described above.
Our goals are two-fold. First, we aim to demonstrate the utility of individual-level covariate adjustment in small CRTs when the binary outcomes are rare, hoping to provide some justification for incorporating baseline individual-level covariates in this particularly challenging scenario. Second, we aim to assess and compare the performances of propensity score weighting and multivariable adjustment models, as well as the associated bias-corrected sandwich estimators in this context. These two aims collectively address an important gap in the existing literature which has primarily focused on evaluating similar methods assuming common binary outcomes and without sufficient attention to individual-level covariate adjustment. %\citep{Li2015,ford2017,li2021sample_bin}.

\subsection{Data-generating mechanism}
We generate CRT data with two parallel arms. Suppose $N$ total clusters are randomized to the two arms under 1:1 randomization. The cluster size for each cluster is drawn from a $Poisson(m)$ distribution, where $m$ is the mean cluster size; the exact number of subjects in cluster $i$ is denoted $m_i$. The outcome ICC under the latent response formulation, $\rho_{Logit}$,  reflects similarity among people in the same cluster \citep{Li2017}. This parameter will be set at values according to the range reported in practice \citep{murray2004}. Defining $P_1$ to be the population incidence under treatment and $P_0$ to be the population incidence under the control, we consider two levels of the outcome incidence---low incidence ($P_1 \approx 0.05, P_0 \approx 0.10$) and very low incidence ($P_1 \approx 0.025, P_0 \approx 0.05$). 

The general combinations of simulation parameters that we consider are summarized as follows: (i) Number of total clusters: $N=\{6, 10, 20, 30\}$; (ii) Mean cluster size: $m=\{100, 30\}$; (iii) ICC on the latent scale: $\rho_{Logit} = \{0.001, 0.01\}$; (iv) Number of covariates: $P=\{6, 15\}$; (v) Incidence levels: low and very low. For each simulation setting, 1000 data sets were generated, and the Monte Carlo errors will be described in Section \ref{sec:perform}. 

For the outcome generating mechanism, we consider four parametric models. First, we simulate covariates from a standard normal distribution, $X^{(p)} \sim N(0,1)$, $p=1,\ldots,P$. The first two outcome models assume a constant additive treatment effect with constant covariate effects on the logistic scale--that is, there are no interactions among variables and no treatment effect heterogeneity explained by covariates. Specifically, the latent continuous outcome for the $j$th subject in the $i$th cluster is
\begin{equation*}
    Y_{ij}^c=\beta_0 + \beta_1 \sum_{p=1}^{P/3} X_{ij}^{(p)} + \beta_2 \sum_{p=P/3+1}^{2P/3} X_{ij}^{(p)} + \beta_3 \sum_{p=2P/3+1}^P X_{ij}^{(p)} + \beta_Z Z_i +u_i + \epsilon_{ij}
\end{equation*}
where $u_i \sim N(0, \sigma_u^2)$ is the random effect and $\epsilon_{ij}$ is assumed to follow the standard logistic distribution with mean $0$ and variance $\sigma_{\epsilon}^2={\pi^2}/{3}$. Then $\sigma_u^2 = {\rho_{Logit}}/({1-\rho_{Logit}})\times ({\pi^2}/{3})$ according to the latent response definition of binary ICC. The binary outcome is obtained by dichotomizing $Y_{ij}^c$: $Y_{ij} \sim Bernoulli\{expit(Y_{ij}^c)\}$.  We consider two sets of fixed coefficients $\beta_1,\beta_2,\beta_3$ for the covariates $X^{(1)},\ldots,X^{(P)}$, which correspond to varying strengths of covariate-outcome associations, and $\beta_Z$ is the main effect of the treatment.
\begin{enumerate}[label=(\roman*)]
\itemsep 0em 
    \item Outcome Generating Model 1: $\beta_1 = 0, \beta_2 = 0.4, \beta_3 = 0.8$. In this model, some of the covariates are not related to the outcome while the others are weakly correlated with the outcome.
    \item Outcome Generating Model 2: $\beta_1 = 0.8, \beta_2 = 1.6, \beta_3 = 2.4$. With this model, all covariates are prognostic and some strongly correlated with the outcome. 
\end{enumerate}
Next, we consider more complex outcome generating models that incorporate nonlinearity and treatment effect heterogeneity (interaction between treatment and covariates). \textcolor{black}{For these models, we only consider six ($P=6$) covariates, which are simulated from a multivariable model with mean vector $\bzero$ and covariance matrix for which diagonal elements are $1$ and off diagonal elements are $0.1$, reflecting weak correlations among the covariates.} 
\begin{enumerate}[resume,label=(\roman*)]
\itemsep 0em 
    \item Outcome Generating Model 3: $Y_{ij}^c = \beta_0 - \displaystyle\frac{3}{1+\exp\left\{-6(X_{ij}^{(1)}+X_{ij}^{(2)}+X_{ij}^{(3)}+X_{ij}^{(4)})\right\}} +\frac{1}{2}(X_{ij}^{(5)}+X_{ij}^{(6)})+2X_{ij}^{(5)}X_{ij}^{(6)}+1.8 (X_{ij}^{(3)}+X_{ij}^{(4)}) Z_{i} - \frac{2}{1+\exp\left\{-4(X_{ij}^{(5)}+X_{ij}^{(6)})\right\}} Z_{i} + \beta_Z Z_{i} +u_i +\epsilon_{ij}$
    \item Outcome Generating Model 4: $Y_{ij}^c = \beta_0-\displaystyle\frac{3}{2(1+\exp\left\{-4(X_{ij}^{(1)}+X_{ij}^{(2)}))\right\}} + 2 sin (X_{ij}^{(3)}+X_{ij}^{(4)}) + 1.8 (X_{ij}^{(1)}X_{ij}^{(3)} + X_{ij}^{(2)}X_{ij}^{(4)}) +X_{ij}^{(5)}+X_{ij}^{(6)}- 1.5 X_{ij}^{(5)} X_{ij}^{(6)}-1.5(X_{ij}^{(3)}+X_{ij}^{(4)})Z_i+\frac{2}{1+\exp\left\{-2(X_{ij}^{(5)}+X_{ij}^{(6)})\right\}}Z_i + \beta_Z Z_i +u_i +\epsilon_{ij}$
\end{enumerate}
Further, the parameters $\beta_0$ and $\beta_Z$ for each combination of generating model, incidence level, and the number of covariates are set at values that give the desired incidences. These more granular considerations on parameter specification are summarized in Table \ref{tab:truth}.

\begin{table}[htbp]
\caption{Specification of key parameters in the outcome generating models.}
\label{tab:truth}
\begin{tabular}{ccccccc}
\toprule
Generating model & Incidence level & \begin{tabular}[c]{@{}c@{}}Number of \\ covariates ($P$)\end{tabular} & $\beta_0$ & $\beta_Z$ & $(P_1,P_0)$     & $\Delta$ \\ \midrule
Outcome 1        & low             & 6                                                                   & -3.6      & -1.2       & (0.0455, 0.0987)  & -0.8317   \\
Outcome 1        & very low        & 6                                                                   & -4.7      & -1.1       & (0.0224, 0.0490)  & -0.8103   \\
Outcome 1        & low             & 15                                                                  & -4.2      & -1.2       & (0.0484, 0.0954)  & -0.7292   \\
Outcome 1        & very low        & 15                                                                  & -5.4      & -1.2       & (0.0221, 0.0486)  & -0.8155   \\ \midrule
Outcome 2        & low             & 6                                                                   & -6.4      & -1.8       & (0.0490, 0.0974)  & -0.7392   \\
Outcome 2        & very low        & 6                                                                   & -8.1      & -1.7       & (0.0240, 0.0507) & -0.7756   \\
Outcome 2        & low             & 15                                                                  & -9.0      & -2.4       & (0.0559, 0.1045)  & -0.6785   \\
Outcome 2        & very low        & 15                                                                  & -11.8     & -2.2       & (0.0254, 0.0499)  & -0.7007   \\ \midrule
Outcome 3        & low             & 6                                                                   & -4.8     &  -2.8     & (0.0498, 0.0988)  & -0.7380   \\
Outcome 3        & very low        & 6                                                                   &  -6.6     & -4.2      & (0.0253, 0.0511) & -0.7298  \\ \midrule
Outcome 4        & low   & 6   & -4.9 & -3.0 & (0.0504, 0.1004) & -0.7433 \\
Outcome 4        & very low & 6 & -6.6 & -3.2 & (0.0246, 0.0490) & -0.7144 \\ \bottomrule
\end{tabular}
\end{table}

\subsection{Estimands}
To address the issue that regression coefficients of different modeling strategies can correspond to different parameters, we have articulated a common, nonparametric causal estimand of interest in Section \ref{sec:estimand}---the participant-average treatment effect on the log odds ratio scale. Note that the main effect of treatment in the data generating model has a conditional interpretation due to other covariates. The ``true" participant-average treatment effect $\Delta$ for each setting is based on the log odds ratio calculated from a simulated population with $N=5000, m=100$. Specifically, each individual has two potential outcomes, which are obtained by taking draws from the Bernoulli distribution with probability given by the outcome generating model under $Z=1$ and $Z=0$, respectively. From these, the population incidences under treatment and under control are obtained, so that \textcolor{black}{the} true parameter value is $\Delta \approx log\{P_1(1-P_0)\}-log\{P_0(1-P_1)\}$, where $P_1$ is the large sample (simulated population) incidence under treatment and $P_0$ is the large sample (simulated population) incidence under control.

\subsection{Methods: analytical strategies with and without baseline individual-level covariates}
We consider two modeling approaches for obtaining propensity scores indicating the estimated probability of being in the treatment group conditional on the subject's baseline covariates. The first is the multivariable logistic model that regresses the cluster-level treatment variable on the main effects of the covariates. We next employ a more flexible BART model with a probit link \citep{chipman2010}, to test if a more flexible propensity score model can potentially adjust for chance imbalances on higher moments of the covariates beyond the mean. Specifically, we intend to test whether using BART for estimating propensity scores provides improvements over traditional parametric propensity scores for individual-level covariate adjustment in CRTs and to identify settings in which that is the case. Once the individual propensity score values are estimated, we construct two types of weight matrices based on IPW and OW. We consider six different models and evaluate their respective performances.
\begin{description}
\itemsep 0em 
    \item[Crude] The crude model is our reference model in which there is no adjustment for individual-level covariates, and only the treatment indicator is included. The model is $logit\{P(Y_{ij}=1)\} = \beta_0 + \beta_Z Z_i$.
    \item[Multi] The multivariable model involves individual-level covariate adjustment and includes the main effects of the covariates along with the treatment indicator, given by $logit\{P(Y_{ij}=1)\} = \beta_0 + \sum_{p=1}^P \beta_p X_{ij}^{(p)} + \beta_Z Z_i$.
    \item[IPW-Logit, IPW-BART, OW-Logit, OW-BART] The individual-level propensity scores are estimated with logistic regression and BART, inverse probability or overlap weights based on the estimated propensity scores are calculated, and then the resulting weight matrix $\bcw_i$ is included in the GEE model for estimation and inference. 
\end{description}
For the \textbf{Crude} approach, we consider the robust sandwich variance estimator and its bias-corrected versions as in \citet{liRedden2015}. For the \textbf{Multi} approach, we consider the sandwich variance estimators described in Table \ref{tab:estimators}(b), and for the propensity score weighting approaches, we consider the sandwich variance estimators defined in Table \ref{tab:estimators}(a). We use the {\tt{R}} statistical software to estimate propensity scores and perform regression analysis. Specifically, the {\tt{dbarts}} package is used to implement BART, and {\tt{geepack}} package is used to fit the GEE models. We developed our own code for computing the sandwich variance estimators in Table \ref{tab:estimators}, and the R code is available in the Supplementary Material.

\subsection{Performance measures}\label{sec:perform}
We report five performance metrics for each combination of simulation setting and analytic approach to compare the relative performances of the methods considered. At each replication $r$ we obtain an estimate of the replication-specific participant-average treatment effect, $\hat{\Delta}_r$. Then over the 1000 replications, we can obtain an estimate of the true participant-average treatment effect, $\Delta=\sum_{r=1}^{1000} \hat{\Delta}_r/1000$. Bias is calculated as the mean difference in each estimate and the true effect value, Bias = $\sum_{r=1}^{1000} (\hat{\Delta}_r-\Delta)/1000$. To determine whether individual-level covariate adjustment provides efficiency gains over the unadjusted model, we present the relative efficiency (RE), which is the ratio of the empirical variance of the crude model to the empirical variance of the regression or propensity score weighting approaches. Further, for each replication, we construct a 95\% normality-based confidence interval. Then the coverage (CVG) using standard error estimates from the robust and bias-corrected estimators is obtained as the proportion of intervals across the replications that includes the true estimand value, $\Delta$. Based on the binomial model with 1000 replications, we consider the coverage between $[93.6\%,96.4\%]$ to be nominal, and in general, higher coverage represents conservative performance which is typically more tolerable than lower coverage (as a reflection of the sandwich variance estimator being anti-conservative). Here, we consider 1000 simulations sufficient for evaluation of coverage because the associated Monte Carlo standard error ranges from $0.7\%$ (when the empirical coverage is $95\%$) to $1.3\%$ (when the empirical coverage is as low as $80\%$) \citep{morris2019}. Lastly, to get a sense of how often separation issues arise for each model, we report the non-convergence proportion (Non-Con), which refers to the proportion of replications that resulted in an error when fitting the model; this metric is particularly relevant as we assume a low incidence binary outcome, and success in model fitting represents a common practical issue for analyzing such data.

\section{Simulation Results} \label{sec:SimResults}

Performance measures for each combination of GEE model, variance estimator, and trial parameters are provided in detail in the supplementary tables. To keep the main illustration simple and concise, we focus on the settings with an average cluster size of 100 and ICC of $0.01$ since the patterns and results do not vary much for the remaining combinations of simulation parameters.

\begin{figure}[htbp]
    \centering
    \includegraphics[width=10cm]{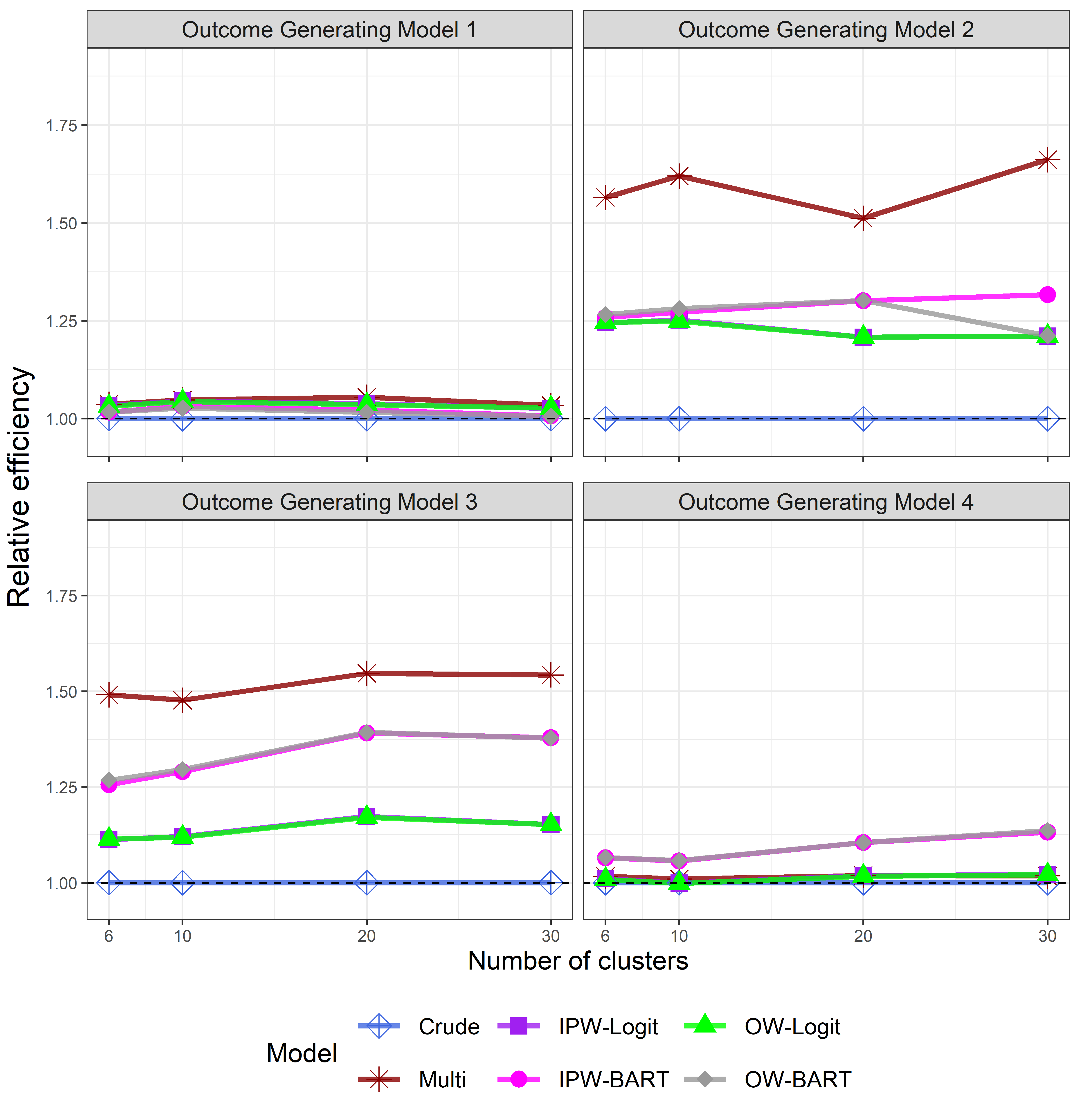}
    \caption{Measures of relative efficiency for 4 different simulation settings with average cluster size of 100 and ICC of 0.01 and assuming low incidence. }
    \label{fig:RE}
\end{figure}

\subsection{Outcome Generating Model 1: Additive effect model that includes weakly prognostic and non-prognostic variables}

\textcolor{black}{\emph{\underline{Efficiency}}:} The relative efficiency of the covariate adjustment methods as compared to that of the crude model for data generated using Outcome Generating Model 1 are close to 1 across the number of clusters considered (Figure \ref{fig:RE} and Supplementary Figure 1). This is also the case for the scenarios with other values of average cluster size ($m$), outcome ICC ($\rho_{Logit}$), and total number of covariates ($P$). Thus, in this setting, there is limited efficiency gain and, in some cases, slightly less efficiency from covariate adjustment if the included variables are unrelated or weakly related to the outcome, but the number of clusters is limited. 

\begin{figure}[htbp]
    \centering
    \includegraphics[width=10cm]{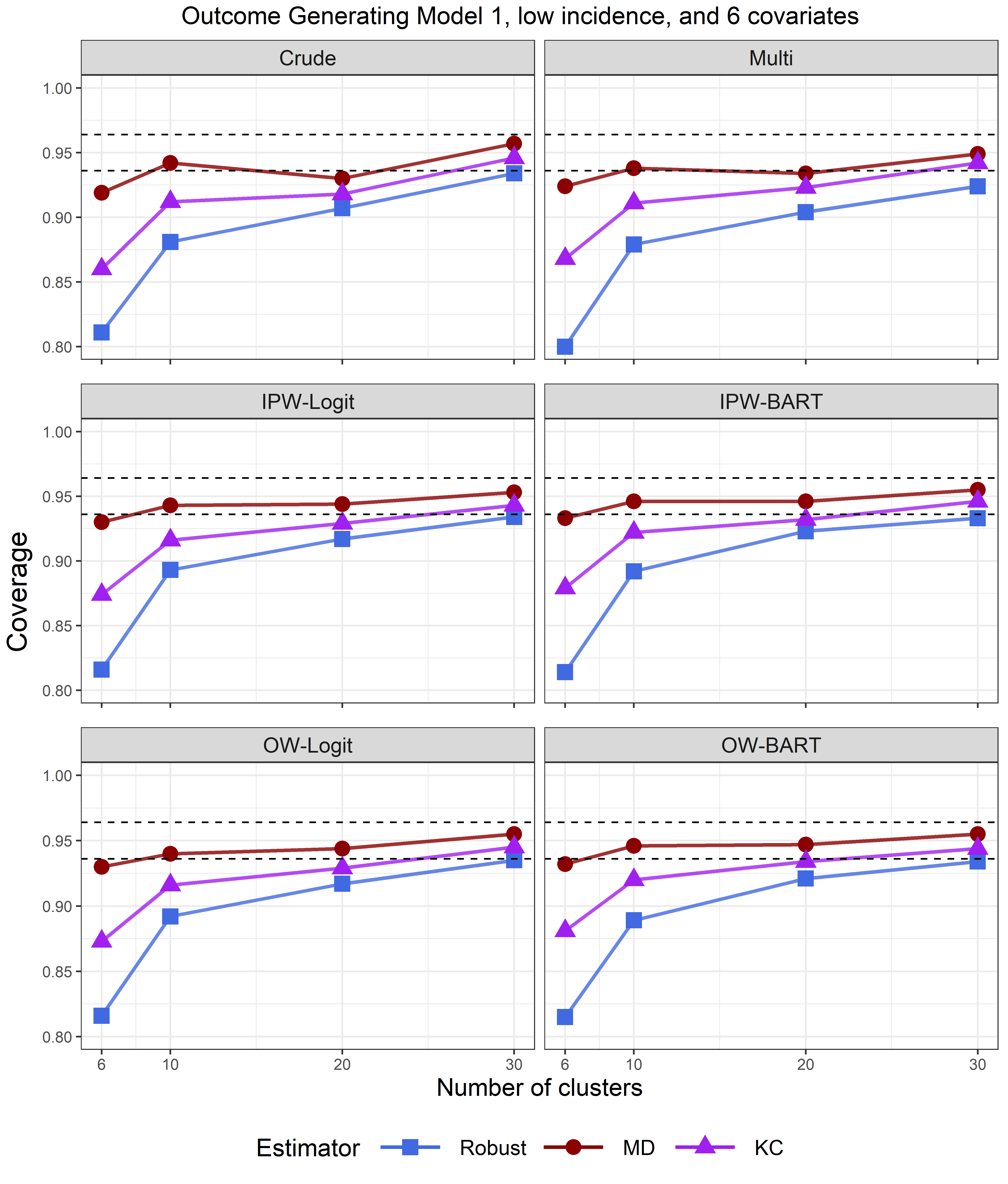}
    \caption{Measures of coverage for simulations based on the Outcome Generating Model 1, low outcome incidences, 6 covariates, average cluster size of 100, and latent ICC of 0.01. Values that are between the dashed lines (at $93.6\%$ and $96.4\%$) are considered nominal.}
    \label{fig:Out1LowCov6m100ICC.01}
\end{figure}

\textcolor{black}{\emph{\underline{Coverage of confidence interval estimator}}:} Figure \ref{fig:Out1LowCov6m100ICC.01} and Supplementary Figure 2 present the coverage rates from 95\% confidence intervals for varying total number of clusters under Outcome Generating Model 1 and assumed low and very low incidences, respectively. For the crude and multivariable models, most of the estimators result in under-coverage for fewer than 30 clusters with the MD bias-corrected sandwich variance estimator giving closest to nominal. Under IPW and OW, the MD bias-corrected sandwich variance estimator generally provides nominal coverage except for the slight undercoverage observed when the number of clusters $N=6$ with logistic propensity score models. 
%Under IPW, the MD bias-corrected sandwich variance estimator also provides nominal coverage with slight undercoverage for the smallest number of clusters, $N=6$. Under OW, the MD bias-corrected sandwich variance estimator provides nominal coverage while the other bias-corrected estimators only reached nominal coverage at $N=30$. 
Overall, even with $N=30$ clusters, the uncorrected robust sandwich estimator may still result in under-coverage, while both bias-corrected estimators give similar coverage rates that are at the nominal level, regardless of whether \textcolor{black}{individual-level} covariate adjustment is considered.

\textcolor{black}{\emph{\underline{Convergence}}:} Non-convergence tends not to be much of an issue when outcome variables are simulated under Outcome Generating Model 1, where there are 6 covariates except at very low incidences and $N=6$ and $m=30$. Here, the non-convergence proportion is $14.7\%$ and $16.0\%$ for the multivariable model and $13.6\%$ and $15.1\%$ for the crude and propensity score weighted models at ICC of $0.001$ and $0.01$, respectively (Supplementary Table 2). If $15$ covariates are adjusted for, approximately half of the replications do not converge for the multivariable model at $6$ clusters with $30$ subjects per cluster on average (Supplementary Table 4). However, weighting-based analysis encounters much less non-convergence (Supplementary Tables 3 and 4) and is a practical solution for \textcolor{black}{individual-level} covariate adjustment when multivariable regression fails to provide a point estimate.

\subsection{Outcome Generating Model 2: Additive effect model that includes variables that are strongly correlated with the outcome}

\textcolor{black}{\emph{\underline{Efficiency}}:} 
Adjustment for covariates that have a large effect on the outcome may substantially increase the efficiency of the model over the crude model (Figure \ref{fig:RE} and Supplementary Figure 1). Further, the multivariable model, when it converges, tends to provide the largest efficiency gain. This is expected because the multivariable model is approximately correctly specified. For the same propensity score estimates (either with logistic regression or BART), IPW and OW provide almost identical estimates of empirical variance. Results are similar for other values of average cluster size ($m$) and outcome ICC ($\rho_{Logit}$). This result is in contrast to those under small individually-randomized trials, where OW dominated IPW in terms of performance \citep{zeng2021propensity}. 

\textcolor{black}{\emph{\underline{Coverage of confidence interval estimator}}:} In Supplementary Figures 3 and 4, the coverage based on 95\% confidence intervals by the number of clusters under Outcome Generating Model 2 and for low and very low incidences, respectively, are given. When the crude and multivariable models were used, all variance estimators result in under-coverage for $N=6, 10$ clusters, although the MD bias-corrected sandwich variance estimator is closest to nominal. Under IPW and OW, the MD bias-corrected sandwich variance estimator consistently leads to nominal coverage. However, once $N=20$ clusters are reached, coverage becomes close to the nominal $0.95$ for all estimators based on propensity score weighting.

\textcolor{black}{\emph{\underline{Convergence}}:} Even at low incidences, the multivariable analysis tends to have larger non-convergence proportions than the crude and propensity score weighted models. Specifically, with $6$ covariates under $N=6$ and $m=30$ at very low incidences, the multivariable model does not converge for $33.3\%$ and $30.9\%$ of the replications when the ICC is $0.001$ and $0.01$, respectively---almost tripling that of the other models (Supplementary Table 6). The problem becomes quite serious when $15$ covariates are adjusted for. The analogous non-convergence proportions are $77.4\%$ and $76.8\%$ when incidences are low (Supplementary Table 7) and $97.6\%$ and $98.6\%$ when incidences are very low (Supplementary Table 8). When $N=10$ and $m=30$, the non-convergence proportions are $79.3\%$ and $82.6\%$ when the ICC is $0.001$ and $0.01$, respectively (Supplementary Table 8). Further, the multivariable model with $15$ covariates is observed to have at least one non-convergence replication even when we have as large as $N=30$ clusters. These findings show that, although the multivariable adjustment GEE often provides the largest efficiency gain, it could exhibit serious non-convergence issues when covariates are strongly prognostic and when the number of clusters is limited. In those scenarios, propensity score weighting, both IPW and OW, becomes a practical solution that offers a moderate precision gain over the unadjusted analysis. When the outcome generating model is relatively simple and additive, machine learning propensity score models do not exhibit any apparent advantage over logistic propensity scores. %and the latter is often sufficient.

\subsection{Outcome Generating Models 3 and 4: Nonlinear covariate-outcome associations with treatment effect heterogeneity explained by individual-level covariates}

\textcolor{black}{\emph{\underline{Efficiency}}:} Under Outcome Generating Model 3, the multivariable analysis gives the largest RE. Further, weighting using BART-estimated propensity scores provides a larger RE than weighting based on logistic-estimated propensity scores, demonstrating that machine learning estimated propensity scores lead to an efficiency advantage over their parametric counterparts when the true outcome model includes complex covariate-outcome associations. Under Outcome Generating Model 4, weighting by BART-estimated propensity scores results in higher efficiency compared to the multivariable analysis. This is somewhat expected because the multivariate analysis does not include correctly specified functional forms of the individual-level covariates. However, this result represents an important piece of evidence that machine learning propensity scores can confer an efficiency advantage in analyzing small CRTs. Further, weighting using BART-estimated propensity scores gives higher RE as the number of clusters increases.

\textcolor{black}{\emph{\underline{Coverage of confidence interval estimator}}:} Under Outcome Generating Model 3 (Supplementary Figures 5 and 6), with propensity score weighting, all the sandwich variance estimators, including the robust estimator, gives coverage that is at least $0.90$ starting with $N=10$ clusters. IPW with BART-estimated propensity scores and OW for both logistic and BART-estimated propensity scores provide coverage close to $0.95$ regardless of the sandwich variance estimator as long as the number of clusters is at least $10$. This is in contrast with the multivariable model, which still results in undercoverage at $N=10$ for the KC bias-corrected sandwich variance estimator. This set of results highlights the important benefit of propensity score weighting as a simple and effective covariate adjustment strategy when the true data generating model is complex. Under Outcome Generating Model 4 (Supplementary Figures 7 and 8), patterns in coverage are generally similar to those observed under Outcome Generating Model 3. 

\textcolor{black}{\emph{\underline{Convergence}}:} For these more complex outcome generating models, the non-convergence proportions from the multivariable analyses remain worse than the weighting approaches at very small number of clusters ($N=6$). The results are summarized in Supplementary Tables 9-12.

\section{Application to the RESTORE Cluster Randomized Trial} \label{sec:Data}

To illustrate the methods for analyzing CRTs with a low incidence outcome, we apply the propensity score weighting and multivariable regression methods to the Randomized Evaluation of Sedation Titration for Respiratory Failure (RESTORE) trial, which was a CRT that took place in $N=31$ U.S. pediatric intensive care units (PICUs). RESTORE compared a nurse-implemented, goal-directed sedation protocol against usual care. The intervention was introduced to 17 PICUs from the Pediatric Acute Lung Injury and Sepsis Investigators (PALISI) Network, while 14 others in this network comprised the control clusters and were given usual care. The total sample size was 2,449 children. Additional details about this study may be found in \citet{curley2015}.

\textcolor{black}{We consider a secondary binary outcome that appeared to have clinically relevant effects---the postextubation stridor outcome ($\hat{P}_1 = .072, \hat{P}_0=.045$). For each of the individual-level covariate adjustment methods, we consider two sets of individual-level covariates corresponding to a small and a large number of variables, all of which are expected to be prognostic, though to varying degrees. The two sets of covariates are: (1) 3 individual-level covariates: age, PRISM III-12 score, and whether baseline POPC score is equal to 1; and (2) 11 individual-level covariates: age, PRISM III-12 score, whether baseline POPC score is equal to 1, pneumonia as primary diagnosis, bronchiolitis as primary diagnosis, acute respiratory failure related to sepsis as primary diagnosis, oxygenation index, prematurity, asthma, cancer (current or previous diagnosis), and intubation at other hospital and transferred to participating PICU.}

The baseline characteristics by treatment arm are summarized in Table \ref{tab:baseline}. The absolute standardized difference (ASD) for each covariate is reported as a measure of imbalance between intervention and control arms. The values for age, PRISM score, bronchiolitis, acute respiratory failure, and asthma are \textcolor{black}{greater} than $0.10$, which is a common threshold for assessing balance \citep{austin2015moving}, suggesting residual imbalance between the treatment arms for these variables. In Supplementary Table 13, we present the propensity score weighted covariate distributions based on the two adjustment sets of interest. With either IPW or OW, covariates that are included in the propensity score model had reduced baseline imbalance based on the weighted ASD, with all of them less than $0.10$. With logistic-estimated propensity scores, OW completely removes the baseline chance imbalance for covariates that were included, due to its exact mean balance property \citep{li2018,zeng2021propensity}. 

\begin{table}[htbp]
%\footnotesize
\centering
\caption{Baseline demographic and clinical characteristics of children who were mechanically ventilated for acute respiratory failure for control and intervention (use of RESTORE protocol) groups. ASD: absolute standardized difference.}
\label{tab:baseline}
\begin{tabular}{cccc}
\toprule
Characteristics & Control & Intervention & ASD \\

 & sample size: $1224$ & sample size: $1225$ &   \\
\midrule
Age (mean (SD)) & 5.21 (5.49) & 4.22 (5.38) & 0.183\\
PRISM score (mean (SD)) & 9.91 (7.50) & 7.93 (7.32) & 0.266\\
Baseline POPC score $= 1$ (\%) & 862 (70.4) & 885 (72.2) & 0.040\\
Pneumonia (\%) & 433 (35.4) & 394 (32.2) & 0.068\\
Bronchiolitis (\%) & 228 (18.6) & 428 (34.9) & 0.375\\
Acute respiratory failure (\%) & 212 (17.3) & 145 (11.8) & 0.156\\
Oxygenation index (mean (SD)) & 8.27 (7.34) & 8.16 (6.85) & 0.015\\
Prematurity (\%) & 175 (14.3) & 194 (15.8) & 0.043\\
Asthma (\%) & 210 (17.2) & 146 (11.9) & 0.149\\
Cancer (\%) & 109 (8.9) & 88 (7.2) & 0.063\\
Transferred (\%) & 306 (25.0) & 334 (27.3) & 0.052\\
\bottomrule
\end{tabular}
\end{table}

\begin{figure}[htbp]
    \centering
    \includegraphics[width=9cm]{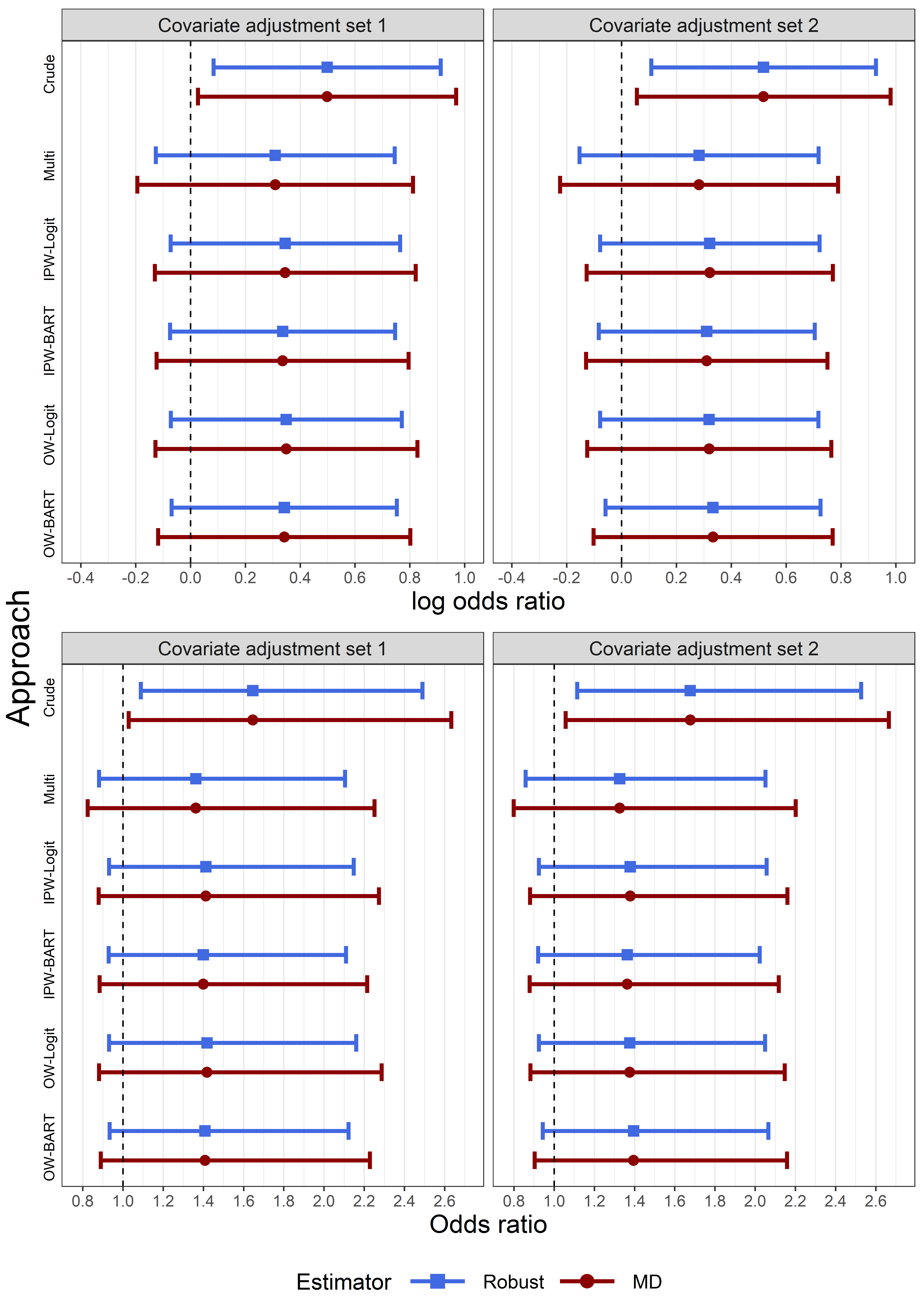}
    \caption{Estimates of participant-average treatment effects and 95\% confidence intervals on the log odds ratio scale (upper panels) and the odds ratio scale (lower panels). Dotted lines at $log(OR) = 0$ and  $OR=1$ indicate no treatment effect.}
    \label{fig:restore}
\end{figure}

Figure \ref{fig:restore} presents the data analysis results, under both sets of adjustment variables, and on both the log OR and OR scales. An overall pattern is that the unadjusted analysis appears to provide a larger participant-average treatment effect compared to any covariate-adjusted analysis. This is possibly due to the moderate chance imbalance in baseline covariates with a limited number of clusters, which may exaggerate the treatment benefit. \textcolor{black}{In our simulation results, the MD bias-corrected sandwich variance estimator frequently leads to nominal coverage for the participant-average treatment effect across methods for individual-level covariate adjustment, and therefore, we only present the confidence intervals based on the MD bias-corrected sandwich variance estimator and the original robust sandwich variance (as a benchmark). On the log OR scale (which corresponds to the estimand in our simulations), the RE values (defined based on the MD bias-corrected sandwich variance estimates) between the covariate-adjusted estimators and the crude estimator are 0.875 (Multivariable), 0.981 (IPW-Logit), 1.048 (IPW-BART), 0.970 (OW-Logit), and 1.049 (OW-BART) using covariate adjustment set 1. The analogous RE values using covariate adjustment set 2 are 0.833 (Multivariable), 1.064 (IPW-Logit), 1.110 (IPW-BART), 1.081 (OW-Logit), and 1.135 (OW-BART). This indicates a slight efficiency advantage of the BART propensity score weighted estimators over the alternatives, and is indicative of possible nonlinear covariate-outcome associations resembling the Outcome Generating Model 4 in our simulations. On the OR scale, individual-level covariate adjustment appears to bring more substantial efficiency gain as shown by the narrower confidence intervals as compared to those obtained based on the crude model. Furthermore, while the new sedation protocol is shown to significantly increase the odds of postextubation stridor, this effect is no longer statistically significant after variables are adjusted for.} 
%However, the potential efficiency benefit of individual-level covariate adjustment was reflected by the narrower confidence intervals after adjustment.}

% However, adjusting for a larger number of covariates tends to increase the efficiency of the estimation and results in more similar estimates across the models that adjusted for covariates. 
% Overall, the multivariable regression analysis leads to results that are similar to those from the propensity score weighting analyses, both in terms of point estimates and 95\% confidence intervals. This might be because the true outcome model is roughly additive, and only a few covariates have weak to moderate prognostic values (akin to our Outcome Generating Model 1). 

\section{Discussion}\label{sec:discussion}

%\textcolor{black}{The present work employed a structured simulation approach to evaluate several individual-level covariate adjustment methods for estimating participant-average treatment effect (defined under a counterfactual outcome framework) in the challenging setting of CRTs with a small number of clusters and a rare binary outcome.} In such settings, 
In this article, we have offered a balanced discussion on the benefits and limitations of propensity score weighting and multivariable adjustment methods, and we identified scenarios under which one of these methods may provide relatively higher statistical efficiency for treatment effect estimation as well as better convergence properties. \textcolor{black}{Furthermore, we have investigated the small-sample inference properties of these estimators with bias-corrected sandwich variance estimators. In what follows, we organize the discussion around three main properties of the competing estimators: efficiency, coverage, and convergence.}

\textcolor{black}{First, regarding efficiency, our results suggest that individual-level covariate adjustment tends to provide better performance than an unadjusted model, especially when the covariates are at least moderately prognostic. This finding parallels those made for individually-randomized trials and reinforces the need for individual-level covariate adjustment in CRTs. However, the optimal individual-level covariate adjustment method can differ based on the underlying data generating process (Figure \ref{fig:RE}). When it is successfully fit, the multivariable model provides the largest efficiency gain when covariates are strongly prognostic of outcome and the model is approximately correctly specified. On the other hand, propensity score weighting may be preferred when data complexities, such as interactions and/or a nonlinear response surface, are expected. In some cases, the increase in efficiency gain from propensity score weighting may be greater from multivariable regression with linear-term adjustment. For weighting estimators, the choice of weights (IPW versus OW) did not have much effect on the performance. Rather, efficiency gains from propensity score weighting varied with the type of model used to estimate the propensity score in some settings. This is in contrast to \citet{zeng2021propensity}, who found OW often dominates IPW in small individually-randomized trials. The likely reason is that when the number of clusters is small, the total sample size remains at least moderate in CRTs. For choices of the propensity score model, BART-estimated propensity scores may provide more efficiency gain as compared to parametric propensity scores, such as logistic regression, when the true outcome data generating model includes possible nonlinear interactions between covariates and treatment. Even though machine learning propensity score models have been shown to provide benefit in analyzing observational studies \citep{lee2010improving}, to the best of our knowledge, this is the first study that demonstrates the utility of a machine learning propensity score model for covariate adjustment in CRTs. Of note, even though BART is inherently Bayesian, we have integrated its posterior mean estimates for the propensity scores into the GEE estimators and have pursued a final frequentist estimator. The consideration of this approach is based on practical utility, and a more formal Bayesian approach is left for future work.}
%rather than relying on theoretical grounds. An alternative approach is to consider an approximate Bayesian inference through Bayesian bootstrap \citep{saarela2015bayesian}, which remains to be explored in CRT applications.}

\textcolor{black}{Second, regarding coverage, we found that the MD bias-corrected sandwich variance estimator frequently provides nominal coverage for individual-level covariate-adjusted estimation of the participant-average treatment effect in CRTs. Although the MD bias-corrected sandwich variance estimator has been shown to be over conservative in previous comparative studies with no or few covariates, it gave approximate $95\%$ coverage in our case, especially when there are very few clusters. We do acknowledge that, however, our findings are based on the $z$-confidence interval rather than the $t$-confidence interval, whereas previous studies, for example, have shown the $t$-test coupled with the KC bias-corrected sandwich variance estimator preserved nominal test size \citep{liRedden2015} when the cluster size is not extremely variable. In our work, the basis for choosing the $z$-confidence interval are two-fold. On one hand, we realize that it may not be straightforward to determine the appropriate degrees of freedom for the $t$-confidence interval with propensity score weighting as technically the marginal mean model only includes a treatment indicator, but potentially many covariates can be adjusted for in the propensity score weights. On the other hand, introducing the $t$-confidence interval with a conservative choice of the degrees of freedom (such as the number of clusters minus the number of covariates adjusted for as in \citet{mancl2001}) may lead to lower power compared to the $z$-confidence interval, which is unfavorable in small CRTs. The optimal degrees of freedom under the $t$-distribution for different individual-level covariate-adjusted estimators and the extent to which the $t$-confidence interval would change our recommendations require further research.}

\textcolor{black}{Finally, regarding convergence, we find that if the outcome incidences are under 0.05, propensity score weighting may be preferred over multivariable regression, which is very likely to have separation and convergence issues. This is especially the case when a large number of individual-level covariates are adjusted for, but the number of clusters is limited. Despite the fact that multivariable regression may lead to the largest efficiency gain when the model is close to correctly specified, the choice between multivariable regression and propensity score weighting for individual-level covariate adjustment in small CRTs should also take into account convergence as a practical challenge. Overall, propensity score weighting represents a useful and feasible approach for individual-level covariate adjustment in CRTs with a rare binary outcome and when the number of clusters is small.}

\textcolor{black}{There are three limitations that we plan to address in future work. First, we have only considered the causal odds ratio as a target estimand. However, the methods are readily applicable to estimate the causal risk difference or relative risk, for example, by choosing the the log link \citep{li2021sample_bin} for unadjusted and weighting-based estimators. It would be valuable to compare individual-level covariate adjustment strategies for estimating these two alternative effect measures. Second, we have restricted to rare binary outcomes with an incidence rate ranging from around $2\%$ to $10\%$, which is similar to that in our motivating study; in this case, we have already observed non-trivial non-convergence issues from fitting multivariable regression models with individual-level covariates. There are other cases with extremely rare outcomes, such as those with incidence rate $<1\%$ \citep{westgate2022marginal}. In that case, while we suspect the non-convergence issue for multivariable regression models can only be more frequent, whether individual-level covariate adjustment can lead to efficiency gain is an open question. Third, we have considered the independence working correlation model, and therefore, the recommendations are limited to independence GEE. As we show in Supplementary File Section 1, the choice of independence correlation structure allows us to interpret the treatment effect coefficient as our target causal estimand by giving each individual equal weight \citep{Brennan2022IJE}. Since the exchangeable working correlation model is another popular choice in analyzing CRTs, it would be interesting to investigate to what extent the current findings can be generalizable to the exchangeable working correlation model.}

\vspace{-0.1in}

%\newpage

\begin{acknowledgement}
Research in this article was partially supported by the Patient-Centered Outcomes Research Institute\textsuperscript{\textregistered} (PCORI\textsuperscript{\textregistered} Awards ME-2020C1-19220 to M.O.H. and ME-2020C3-21072 to F.L). M.O.H. is also funded by the United States National Institutes of Health (NIH), National Heart, Lung, and Blood Institute (grant number R00-HL141678). All statements in this report, including its findings and conclusions, are solely those of the authors and do not necessarily represent the views of the NIH or PCORI\textsuperscript{\textregistered} or its Board of Governors or Methodology Committee. The authors thank the Special Issue Editor and two anonymous reviewers for their helpful suggestions that have substantially improved this work. The authors also thank Can Meng for providing computational assistance during the revision of this work.
\end{acknowledgement}

\vspace{0.1in}

\noindent {\bf{Conflict of Interest}}

\noindent {The authors declare no conflict of interest.}

\vspace{0.1in}

\noindent {\bf{Data availability statement}}

\begin{sloppypar}
\noindent {Data from the RESTORE trial was obtained from the United States National Heart, Lung and Blood Institute, Biologic Specimen and Data Repository Information Coordinating Center (BioLINCC)\\ (\textcolor{black}{https://biolincc.nhlbi.nih.gov/home/}). The authors do not have the permission to share this data directly; however, the data is available to researchers upon request and approval from BioLINCC.}
\end{sloppypar}

\bibliographystyle{apalike} 
\bibliography{references}

\begin{thebibliography}{}

\bibitem[Austin and Stuart, 2015]{austin2015moving}
Austin, P.~C. and Stuart, E.~A. (2015).
\newblock Moving towards best practice when using inverse probability of
  treatment weighting (iptw) using the propensity score to estimate causal
  treatment effects in observational studies.
\newblock {\em Statistics in Medicine}, 34(28):3661--3679.

\bibitem[Benkeser et~al., 2021]{benkeser2021}
Benkeser, D., D{\'\i}az, I., Luedtke, A., Segal, J., Scharfstein, D., and
  Rosenblum, M. (2021).
\newblock Improving precision and power in randomized trials for covid-19
  treatments using covariate adjustment, for binary, ordinal, and time-to-event
  outcomes.
\newblock {\em Biometrics}, 77(4):1467--1481.

\bibitem[Chipman et~al., 2010]{chipman2010}
Chipman, H.~A., George, E.~I., and McCulloch, R.~E. (2010).
\newblock {BART}: Bayesian additive regression trees.
\newblock {\em The Annals of Applied Statistics}, 4(1):266--298.

\bibitem[Curley et~al., 2015]{curley2015}
Curley, M.~A., Wypij, D., Watson, R.~S., Grant, M. J.~C., Asaro, L.~A.,
  Cheifetz, I.~M., Dodson, B.~L., Franck, L.~S., Gedeit, R.~G., Angus, D.~C.,
  et~al. (2015).
\newblock Protocolized sedation vs usual care in pediatric patients
  mechanically ventilated for acute respiratory failure: a randomized clinical
  trial.
\newblock {\em JAMA}, 313(4):379--389.

\bibitem[Dorie et~al., 2019]{dorie2019automated}
Dorie, V., Hill, J., Shalit, U., Scott, M., and Cervone, D. (2019).
\newblock Automated versus do-it-yourself methods for causal inference: Lessons
  learned from a data analysis competition.
\newblock {\em Statistical Science}, 34(1):43--68.

\bibitem[Fay and Graubard, 2001]{fay2001}
Fay, M.~P. and Graubard, B.~I. (2001).
\newblock Small-sample adjustments for wald-type tests using sandwich
  estimators.
\newblock {\em Biometrics}, 57(4):1198--1206.

\bibitem[Ford and Westgate, 2017]{ford2017}
Ford, W.~P. and Westgate, P.~M. (2017).
\newblock Improved standard error estimator for maintaining the validity of
  inference in cluster randomized trials with a small number of clusters.
\newblock {\em Biometrical Journal}, 59(3):478--495.

\bibitem[Hill, 2011]{hill2011}
Hill, J.~L. (2011).
\newblock Bayesian nonparametric modeling for causal inference.
\newblock {\em Journal of Computational and Graphical Statistics},
  20(1):217--240.

\bibitem[Ivers et~al., 2011]{ivers2011}
Ivers, N., Taljaard, M., Dixon, S., Bennett, C., McRae, A., Taleban, J., Skea,
  Z., Brehaut, J., Boruch, R., Eccles, M., et~al. (2011).
\newblock Impact of consort extension for cluster randomised trials on quality
  of reporting and study methodology: review of random sample of 300 trials,
  2000-8.
\newblock {\em BMJ}, 343.

\bibitem[Kahan et~al., 2022]{Brennan2022IJE}
Kahan, B., Li, F., Copas, A., and Harhay, M.~O. (2022).
\newblock Estimands in cluster-randomised trials: choosing analyses that answer
  the right question.
\newblock {\em International Journal of Epidemiology}, 00(0):000.

\bibitem[Kauermann and Carroll, 2001]{kauermann2001}
Kauermann, G. and Carroll, R.~J. (2001).
\newblock A note on the efficiency of sandwich covariance matrix estimation.
\newblock {\em Journal of the American Statistical Association},
  96(456):1387--1396.

\bibitem[Lee et~al., 2010]{lee2010improving}
Lee, B.~K., Lessler, J., and Stuart, E.~A. (2010).
\newblock Improving propensity score weighting using machine learning.
\newblock {\em Statistics in Medicine}, 29(3):337--346.

\bibitem[Leyrat et~al., 2013]{leyrat2013propensity}
Leyrat, C., Caille, A., Donner, A., and Giraudeau, B. (2013).
\newblock Propensity scores used for analysis of cluster randomized trials with
  selection bias: a simulation study.
\newblock {\em Statistics in Medicine}, 32(19):3357--3372.

\bibitem[Leyrat et~al., 2014]{leyrat2014propensity}
Leyrat, C., Caille, A., Donner, A., and Giraudeau, B. (2014).
\newblock Propensity score methods for estimating relative risks in cluster
  randomized trials with low-incidence binary outcomes and selection bias.
\newblock {\em Statistics in Medicine}, 33(20):3556--3575.

\bibitem[Leyrat et~al., 2018]{leyrat2018}
Leyrat, C., Morgan, K.~E., Leurent, B., and Kahan, B.~C. (2018).
\newblock Cluster randomized trials with a small number of clusters: which
  analyses should be used?
\newblock {\em International Journal of Epidemiology}, 47(1):321--331.

\bibitem[Li et~al., 2015]{Li2015}
Li, F., Lokhnygina, Y., Murray, D.~M., Heagerty, P.~J., and Delong, E.~R.
  (2015).
\newblock {An evaluation of constrained randomization for the design and
  analysis of group-randomized trials}.
\newblock {\em Statistics in Medicine}, 35(10):1565--1579.

\bibitem[Li et~al., 2018a]{li2018}
Li, F., Morgan, K.~L., and Zaslavsky, A.~M. (2018a).
\newblock Balancing covariates via propensity score weighting.
\newblock {\em Journal of the American Statistical Association},
  113(521):390--400.

\bibitem[Li et~al., 2021]{li2021clarifying}
Li, F., Tian, Z., Bobb, J., Papadogeorgou, G., and Li, F. (2021).
\newblock Clarifying selection bias in cluster randomized trials.
\newblock {\em Clinical Trials}, page 17407745211056875.

\bibitem[Li and Tong, 2021a]{li2021sample}
Li, F. and Tong, G. (2021a).
\newblock Sample size and power considerations for cluster randomized trials
  with count outcomes subject to right truncation.
\newblock {\em Biometrical Journal}, 63(5):1052--1071.

\bibitem[Li and Tong, 2021b]{li2021sample_bin}
Li, F. and Tong, G. (2021b).
\newblock Sample size estimation for modified poisson analysis of cluster
  randomized trials with a binary outcome.
\newblock {\em Statistical Methods in Medical Research}, 30(5):1288--1305.

\bibitem[Li et~al., 2017]{Li2017}
Li, F., Turner, E.~L., Heagerty, P.~J., Murray, D.~M., Vollmer, W.~M., and
  Delong, E.~R. (2017).
\newblock {An evaluation of constrained randomization for the design and
  analysis of group-randomized trials with binary outcomes}.
\newblock {\em Statistics in Medicine}, 36:3791--3806.

\bibitem[Li et~al., 2018b]{li2018sample}
Li, F., Turner, E.~L., and Preisser, J.~S. (2018b).
\newblock Sample size determination for gee analyses of stepped wedge cluster
  randomized trials.
\newblock {\em Biometrics}, 74(4):1450--1458.

\bibitem[Li and Redden, 2015]{liRedden2015}
Li, P. and Redden, D.~T. (2015).
\newblock Small sample performance of bias-corrected sandwich estimators for
  cluster-randomized trials with binary outcomes.
\newblock {\em Statistics in Medicine}, 34(2):281--296.

\bibitem[Liang and Zeger, 1986]{Liang1986}
Liang, K.-Y. and Zeger, S.~L. (1986).
\newblock {Longitudinal data analysis using generalized linear models}.
\newblock {\em Biometrika}, 73(1):13--22.

\bibitem[Lu et~al., 2007]{lu2007}
Lu, B., Preisser, J.~S., Qaqish, B.~F., Suchindran, C., Bangdiwala, S.~I., and
  Wolfson, M. (2007).
\newblock A comparison of two bias-corrected covariance estimators for
  generalized estimating equations.
\newblock {\em Biometrics}, 63(3):935--941.

\bibitem[Mancl and DeRouen, 2001]{mancl2001}
Mancl, L.~A. and DeRouen, T.~A. (2001).
\newblock A covariance estimator for gee with improved small-sample properties.
\newblock {\em Biometrics}, 57(1):126--134.

\bibitem[Morris et~al., 2019]{morris2019}
Morris, T.~P., White, I.~R., and Crowther, M.~J. (2019).
\newblock Using simulation studies to evaluate statistical methods.
\newblock {\em Statistics in Medicine}, 38(11):2074--2102.

\bibitem[Murray, 1998]{Murray1998}
Murray, D.~M. (1998).
\newblock {\em {Design and Analysis of Group-Randomized Trials}}.
\newblock Oxford University Press, New York, NY.

\bibitem[Murray et~al., 2004]{murray2004}
Murray, D.~M., Varnell, S.~P., and Blitstein, J.~L. (2004).
\newblock Design and analysis of group-randomized trials: a review of recent
  methodological developments.
\newblock {\em American Journal of Public Health}, 94(3):423--432.

\bibitem[Preisser et~al., 2003]{Preisser2003}
Preisser, J.~S., Young, M.~L., Zaccaro, D.~J., and Wolfson, M. (2003).
\newblock {An integrated population-averaged approach to the design, analysis
  and sample size determination of cluster-unit trials}.
\newblock {\em Statistics in Medicine}, 22(8):1235--1254.

\bibitem[Rosenbaum and Rubin, 1983]{rosenbaum1983central}
Rosenbaum, P.~R. and Rubin, D.~B. (1983).
\newblock The central role of the propensity score in observational studies for
  causal effects.
\newblock {\em Biometrika}, 70(1):41--55.

\bibitem[Scott et~al., 2017]{scott2017finite}
Scott, J.~M., deCamp, A., Juraska, M., Fay, M.~P., and Gilbert, P.~B. (2017).
\newblock Finite-sample corrected generalized estimating equation of population
  average treatment effects in stepped wedge cluster randomized trials.
\newblock {\em Statistical methods in medical research}, 26(2):583--597.

\bibitem[Su and Ding, 2021]{su2021model}
Su, F. and Ding, P. (2021).
\newblock Model-assisted analyses of cluster-randomized experiments.
\newblock {\em Journal of the Royal Statistical Society: Series B (Statistical
  Methodology)}.

\bibitem[Turner et~al., 2017a]{turner2017a}
Turner, E.~L., Li, F., Gallis, J.~A., Prague, M., and Murray, D.~M. (2017a).
\newblock Review of recent methodological developments in group-randomized
  trials: part 1—design.
\newblock {\em American Journal of Public Health}, 107(6):907--915.

\bibitem[Turner et~al., 2017b]{turner2017b}
Turner, E.~L., Prague, M., Gallis, J.~A., Li, F., and Murray, D.~M. (2017b).
\newblock Review of recent methodological developments in group-randomized
  trials: part 2—analysis.
\newblock {\em American Journal of Public Health}, 107(7):1078--1086.

\bibitem[Wang et~al., 2022a]{wang2022two}
Wang, X., Turner, E.~L., Li, F., Wang, R., Moyer, J., Cook, A.~J., Murray,
  D.~M., and Heagerty, P.~J. (2022a).
\newblock Two weights make a wrong: Cluster randomized trials with variable
  cluster sizes and heterogeneous treatment effects.
\newblock {\em Contemporary Clinical Trials}, 114:106702.

\bibitem[Wang et~al., 2022b]{wang2022power}
Wang, X., Turner, E.~L., Preisser, J.~S., and Li, F. (2022b).
\newblock Power considerations for generalized estimating equations analyses of
  four-level cluster randomized trials.
\newblock {\em Biometrical Journal}, 64(4):663--680.

\bibitem[Watson et~al., 2021]{watson2021design}
Watson, S.~I., Girling, A., and Hemming, K. (2021).
\newblock Design and analysis of three-arm parallel cluster randomized trials
  with small numbers of clusters.
\newblock {\em Statistics in Medicine}, 40(5):1133--1146.

\bibitem[Westgate, 2013]{westgate2013small}
Westgate, P.~M. (2013).
\newblock On small-sample inference in group randomized trials with binary
  outcomes and cluster-level covariates.
\newblock {\em Biometrical Journal}, 55(5):789--806.

\bibitem[Westgate et~al., 2022]{westgate2022marginal}
Westgate, P.~M., Cheng, D.~M., Feaster, D.~J., Fern{\'a}ndez, S., Shoben,
  A.~B., and Vandergrift, N. (2022).
\newblock Marginal modeling in community randomized trials with rare events:
  Utilization of the negative binomial regression model.
\newblock {\em Clinical Trials}, 19(2):162--171.

\bibitem[Westreich et~al., 2010]{westreich2010}
Westreich, D., Lessler, J., and Funk, M.~J. (2010).
\newblock Propensity score estimation: neural networks, support vector
  machines, decision trees (cart), and meta-classifiers as alternatives to
  logistic regression.
\newblock {\em Journal of Clinical Epidemiology}, 63(8):826--833.

\bibitem[Williamson et~al., 2014]{williamson2014variance}
Williamson, E.~J., Forbes, A., and White, I.~R. (2014).
\newblock Variance reduction in randomised trials by inverse probability
  weighting using the propensity score.
\newblock {\em Statistics in Medicine}, 33(5):721--737.

\bibitem[Zeng et~al., 2021]{zeng2021propensity}
Zeng, S., Li, F., Wang, R., and Li, F. (2021).
\newblock Propensity score weighting for covariate adjustment in randomized
  clinical trials.
\newblock {\em Statistics in Medicine}, 40(4):842--858.

\bibitem[Zhou et~al., 2022]{zhou2021constrained}
Zhou, Y., Turner, E.~L., Simmons, R.~A., and Li, F. (2022).
\newblock Constrained randomization and statistical inference for multi-arm
  parallel cluster randomized controlled trials.
\newblock {\em Statistics in Medicine}.

\bibitem[Zhu et~al., 2021]{zhu2021}
Zhu, Y., Hubbard, R.~A., Chubak, J., Roy, J., and Mitra, N. (2021).
\newblock Core concepts in pharmacoepidemiology: Violations of the positivity
  assumption in the causal analysis of observational data: Consequences and
  statistical approaches.
\newblock {\em Pharmacoepidemiology and Drug Safety}, 30(11):1471--1485.

\end{thebibliography}

% \textit{Please insert appendices before the references.}

%\newpage
%\phantom{aaaa}
\end{document}

\maketitle

\section{Estimation of participant-average causal odds ratio under an independence GEE without additional covariates}

Here we briefly outline the key steps to demonstrate that the GEE estimator adjusting for the cluster-level treatment indicator under the independence working correlation provides an (asymptotically) unbiased quantification of the participant-average log odds ratio. Resuming the notation in the main paper, $Z_i$ denotes the treatment status of cluster $i$ and $Y_{ij}$ is the outcome for subject $j$ in cluster $i$ for a CRT with $N$ total clusters with each cluster containing $m_i$ subjects for $i=1,...,N$. Let $\beta_0$ denote the intercept and $\beta_Z$ denote the coefficient for the treatment variable in the GEE model. Assuming an independence working correlation matrix, the estimating equations to be solved are
\begin{equation*}
    \sum_{i=1}^N \begin{bmatrix} 1 \\ Z_i \end{bmatrix} \sum_{j=1}^{m_i} \left(Y_{ij}-\frac{exp(\beta_0+\beta_ZZ_i)}{1+exp(\beta_0+\beta_ZZ_i)} \right) = \bzero
\end{equation*}
Since the right hand side is the $\bzero$ vector, the elements on the left hand side must all be equal to 0. Thus, the solution to the above equation must satisfy
\begin{align*}
    0 &= \sum_{i=1}^N\sum_{j=1}^{m_i} \left(Y_{ij}-\frac{exp(\beta_0+\beta_ZZ_i)}{1+exp(\beta_0+\beta_ZZ_i)} \right) \\
    &= \sum_{i=1}^N\sum_{j=1}^{m_i}  Z_i \left(Y_{ij}-\frac{exp(\beta_0+\beta_Z)}{1+exp(\beta_0+\beta_Z)} \right) + \sum_{i=1}^N\sum_{j=1}^{m_i}  (1-Z_i)  \left(Y_{ij}-\frac{exp(\beta_0)}{1+exp(\beta_0)} \right) 
\end{align*}
and \begin{equation*}
    \sum_{i=1}^N\sum_{j=1}^{m_i}  Z_i \left(Y_{ij}-\frac{exp(\beta_0+\beta_Z)}{1+exp(\beta_0+\beta_Z)} \right)=0
\end{equation*}
Then we know that \begin{equation*}
    \sum_{i=1}^N\sum_{j=1}^{m_i}  (1-Z_i) \left(Y_{ij}-\frac{exp(\beta_0)}{1+exp(\beta_0)} \right)=0.
\end{equation*}
Rearranging terms, we have
\begin{align*}
    \frac{exp(\beta_0)}{1+exp(\beta_0)} = \frac{\sum_{i=1}^N\sum_{j=1}^{m_i}  (1-Z_i)Y_{ij}}{\sum_{i=1}^N (1-Z_i)m_i},~~~~\qquad
    \frac{exp(\beta_0+\beta_Z)}{1+exp(\beta_0+\beta_Z)} = \frac{\sum_{i=1}^N\sum_{j=1}^{m_i}  Z_iY_{ij}}{\sum_{i=1}^N Z_im_i}
\end{align*}

so that the GEE estimators can be expressed as
\begin{align*}
    \hat{\beta}_0 &= logit\left\{\frac{\sum_{i=1}^N\sum_{j=1}^{m_i} (1-Z_i)Y_{ij}}{\sum_{i=1}^N (1-Z_i)m_i}\right\} \\
    \hat{\beta}_Z &= logit\left\{\frac{\sum_{i=1}^N\sum_{j=1}^{m_i}  Z_i Y_{ij}}{\sum_{i=1}^N Z_i m_i}\right\} -  logit\left\{\frac{\sum_{i=1}^N\sum_{j=1}^{m_i} (1-Z_i)Y_{ij}}{\sum_{i=1}^N (1-Z_i)m_i}\right\} = logit(\hat{P}_1) - logit(\hat{P}_0)
\end{align*}
Thus, we obtain the independence GEE estimator as $exp(\hat{\beta}_Z) = \{\hat{P}_1 (1-\hat{P_0})\}/\{(1-\hat{P}_1)\hat{P_0}\}$. Furthermore, we can show that 
\begin{align*}
\hat{P}_0=&\frac{N^{-1}\sum_{i=1}^N\sum_{j=1}^{m_i} (1-Z_i)Y_{ij}}{N^{-1}\sum_{i=1}^N (1-Z_i)m_i}\\
\overset{p}{\to}&\frac{E\left(\sum_{j=1}^{m_i} (1-Z_i)Y_{ij}\right)}{E\left((1-Z_i)m_i\right)}=\frac{E\left(\sum_{j=1}^{m_i} (1-Z_i)Y_{ij}(0)\right)}{E\left((1-Z_i)m_i\right)}=\frac{E(1-Z_i)E\left(\sum_{j=1}^{m_i}Y_{ij}(0)\right)}{E(1-Z_i)E\left(m_i\right)}=\frac{E\left(\sum_{j=1}^{m_i}Y_{ij}(0)\right)}{E\left(m_i\right)},
\end{align*}
where the convergence in probability statement results from an application of the Weak Law of Large Number for independent but non-identically distributed data, the first equality in the second line is due to the consistency assumption that $Y_{ij}=Z_1Y_{ij}(1)+(1-Z_i)Y_{ij}(0)$, and the second equality in that same line is due to cluster-level randomization. Following the exact steps, we observe that
\begin{align*}
\hat{P}_1\overset{p}{\to}&\frac{E\left(\sum_{j=1}^{m_i}Y_{ij}(1)\right)}{E\left(m_i\right)},
\end{align*}
and therefore
\begin{align*}
exp(\hat{\beta}_Z) \overset{p}{\to}     \frac{E\left(\sum_{j=1}^{m_i} Y_{ij}(1)\right) E\left(m_i-\sum_{j=1}^{m_i} Y_{ij}(0)\right)}{E\left(\sum_{j=1}^{m_i} Y_{ij}(0)\right)E\left(m_i-\sum_{j=1}^{m_i} Y_{ij}(1)\right)},
\end{align*}
which is the participant-average treatment effect on the odds ratio scale. And it is immediate that $\hat{\beta}_Z \overset{p}{\to} \Delta$.

Finally, note that $\hat{P}_1$ is the outcome incidence in the treated group and $\hat{P}_0$ is the outcome incidence in the control group. Adding the propensity score weights to the independence GEE estimator does not change the limit of $\hat{\beta}_Z$ due to cluster randomization (as the estimated propensity scores converge to a constant), and therefore $\hat{\beta}_Z$ in a propensity score weighted GEE is also a valid estimator for $\Delta$. In contrast, if the working correlation model is exchangeable, $\hat{\beta}_Z$ may converge to a different target parameter than $\Delta$. For example, the probability limit of the exchangeable correlation estimator $\hat{\beta}_Z$ will involve an unknown intracluster correlation coefficient and will often differ from $\Delta$ unless the cluster sizes $m_i$'s are all equal; see, for example, \citet{wang2022two}.

\bibliographystyle{apalike} 
\bibliography{references}

\section{Derivation of the $\bcm$ matrix for the multivariable regression estimator}

Based on the GEE model fit, the predicted risk for subject $j$ in cluster $i$, if he/she is given treatment is $\hat{P}_{1, ij} = {exp(\bcx_{1, ij}'\hat{\boldsymbol\beta})}/\left\{1+exp(\bcx_{1,ij}'\hat{\boldsymbol\beta})\right\}$, where $\bcx_{1, ij} = [1, \bcx_{ij}', 1]'$. Similarly, the predicted risk for subject $j$ in cluster $i$ under the control is $\hat{P}_{0, ij} = {exp(\bcx_{0, ij}'\hat{\boldsymbol\beta})}/\left\{1+exp(\bcx_{0, ij}'\hat{\boldsymbol\beta})\right\}$
where $\bcx_{0, ij} = [1, \bcx_{ij}', 0]'$. Then a standardization estimator of the participant-average treatment effect is
{\footnotesize\begin{equation*}
    log (\widehat{OR}) = log \left\{ \frac{\left(\sum_{i=1}^N\sum_{j=1}^{m_i} \hat{P}_{1,ij}\right) \left(\sum_{i=1}^N\left(m_i-\sum_{j=1}^{m_i} \hat{P}_{0,ij}\right)\right)}{\left(\sum_{i=1}^N\sum_{j=1}^{m_i} \hat{P}_{0,ij}\right)\left(\sum_{i=1}^N\left(m_i-\sum_{j=1}^{m_i} \hat{P}_{1,ij}\right)\right)}\right\}
\end{equation*}}
Using the delta method, we arrive at $\widehat{Var} (log (\widehat{OR}))= \bcm' Cov(\hat{\boldsymbol{\beta}}) \bcm$, where $\bcm = \partial log (\widehat{OR})/\partial\boldsymbol{\beta}$.
{\footnotesize
\begin{align*}
     \frac{\partial}{\partial\beta} log \left(\sum_{i=1}^N\sum_{j=1}^{m_i} \hat{P}_{1,ij} \right) 
     =& \frac{1}{\sum_{i=1}^N\sum_{j=1}^{m_i} \hat{P}_{1,ij}} \left\{\sum_{i=1}^N\sum_{j=1}^{m_i} \frac{exp(\bcx_{j1}'\hat{\boldsymbol{\beta}})\bcx_{1,ij}(1+exp(\bcx_{1,ij}'\hat{\boldsymbol{\beta}})-exp(\bcx_{1,ij}'\hat{\boldsymbol{\beta}})\bcx_{1,ij}exp(\bcx_{1,ij}'\hat{\boldsymbol{\beta}})}{[1+exp(\bcx_{1,ij}'\hat{\boldsymbol{\beta}})]^2} \right\} \\
     =& \frac{1}{\sum_{i=1}^N\sum_{j=1}^{m_i} \hat{P}_{1,ij}} \left\{\sum_{i=1}^N\sum_{j=1}^{m_i} \frac{exp(\bcx_{1,ij}'\hat{\boldsymbol{\beta}})\bcx_{1,ij}}{[1+exp(\bcx_{1,ij}'\hat{\boldsymbol{\beta}})]^2} \right\} \\
     \frac{\partial}{\partial\boldsymbol{\beta}} log \left(J-\sum_{i=1}^N\sum_{j=1}^{m_i} \hat{P}_{0,ij} \right) 
     &= \frac{-1}{J-\sum_{i=1}^N\sum_{j=1}^{m_i} \hat{P}_{0,ij}} \left\{\sum_{i=1}^N\sum_{j=1}^{m_i} \frac{exp(\bcx_{0,ij}'\hat{\boldsymbol{\beta}})\bcx_{0,ij}(1+exp(\bcx_{0,ij}'\hat{\boldsymbol{\beta}})-exp(\bcx_{0,ij}'\hat{\boldsymbol{\beta}})\bcx_{0,ij}exp(\bcx_{0,ij}'\hat{\boldsymbol{\beta}})}{[1+exp(\bcx_{0,ij}'\hat{\boldsymbol{\beta}})]^2} \right\} \\
     &= -\frac{1}{J-\sum_{i=1}^N\sum_{j=1}^{m_i} \hat{P}_{0,ij}} \left\{\sum_{i=1}^N\sum_{j=1}^{m_i} \frac{exp(\bcx_{0,ij}'\hat{\boldsymbol{\beta}})\bcx_{0,ij}}{[1+exp(\bcx_{0,ij}'\hat{\boldsymbol{\beta}})]^2} \right\} \\
     \frac{\partial}{\partial\boldsymbol{\beta}} log \left(J-\sum_{i=1}^N\sum_{j=1}^{m_i} \hat{P}_{1,ij} \right)
     &= \frac{-1}{J - \sum_{i=1}^N\sum_{j=1}^{m_i} \hat{P}_{1,ij}} \left\{\sum_{i=1}^N\sum_{j=1}^{m_i} \frac{exp(\bcx_{j1}'\hat{\boldsymbol{\beta}})\bcx_{1,ij}(1+exp(\bcx_{1,ij}'\hat{\boldsymbol{\beta}})-exp(\bcx_{1,ij}'\hat{\boldsymbol{\beta}})\bcx_{1,ij}exp(\bcx_{1,ij}'\hat{\boldsymbol{\beta}})}{[1+exp(\bcx_{1,ij}'\hat{\boldsymbol{\beta}})]^2} \right\} \\
     &= -\frac{1}{J - \sum_{i=1}^N\sum_{j=1}^{m_i} \hat{P}_{1,ij}} \left\{\sum_{i=1}^N\sum_{j=1}^{m_i} \frac{exp(\bcx_{1,ij}'\hat{\boldsymbol{\beta}})\bcx_{1,ij}}{[1+exp(\bcx_{1,ij}'\hat{\boldsymbol{\beta}})]^2} \right\}\\
     \frac{\partial}{\partial\boldsymbol{\beta}} log \left(\sum_{i=1}^N\sum_{j=1}^{m_i} \hat{P}_{0,ij} \right) 
     &= \frac{1}{\sum_{i=1}^N\sum_{j=1}^{m_i} \hat{P}_{0,ij}} \left\{\sum_{i=1}^N\sum_{j=1}^{m_i} \frac{exp(\bcx_{0,ij}'\hat{\boldsymbol{\beta}})\bcx_{0,ij}(1+exp(\bcx_{0,ij}'\hat{\boldsymbol{\beta}})-exp(\bcx_{0,ij}'\hat{\boldsymbol{\beta}})\bcx_{0,ij}exp(\bcx_{0,ij}'\hat{\boldsymbol{\beta}})}{[1+exp(\bcx_{0,ij}'\hat{\boldsymbol{\beta}})]^2} \right\} \\
     &= \frac{1}{\sum_{i=1}^N\sum_{j=1}^{m_i} \hat{P}_{0,ij}} \left\{\sum_{i=1}^N\sum_{j=1}^{m_i} \frac{exp(\bcx_{0,ij}'\hat{\boldsymbol{\beta}})\bcx_{0,ij}}{[1+exp(\bcx_{0,ij}'\hat{\boldsymbol{\beta}})]^2} \right\}
 \end{align*}
}

\section{Additional simulation results}

\subsection{Relative efficiency under very low incidence outcome}

\begin{figure}[H]
    \centering
    \includegraphics[width=13cm]{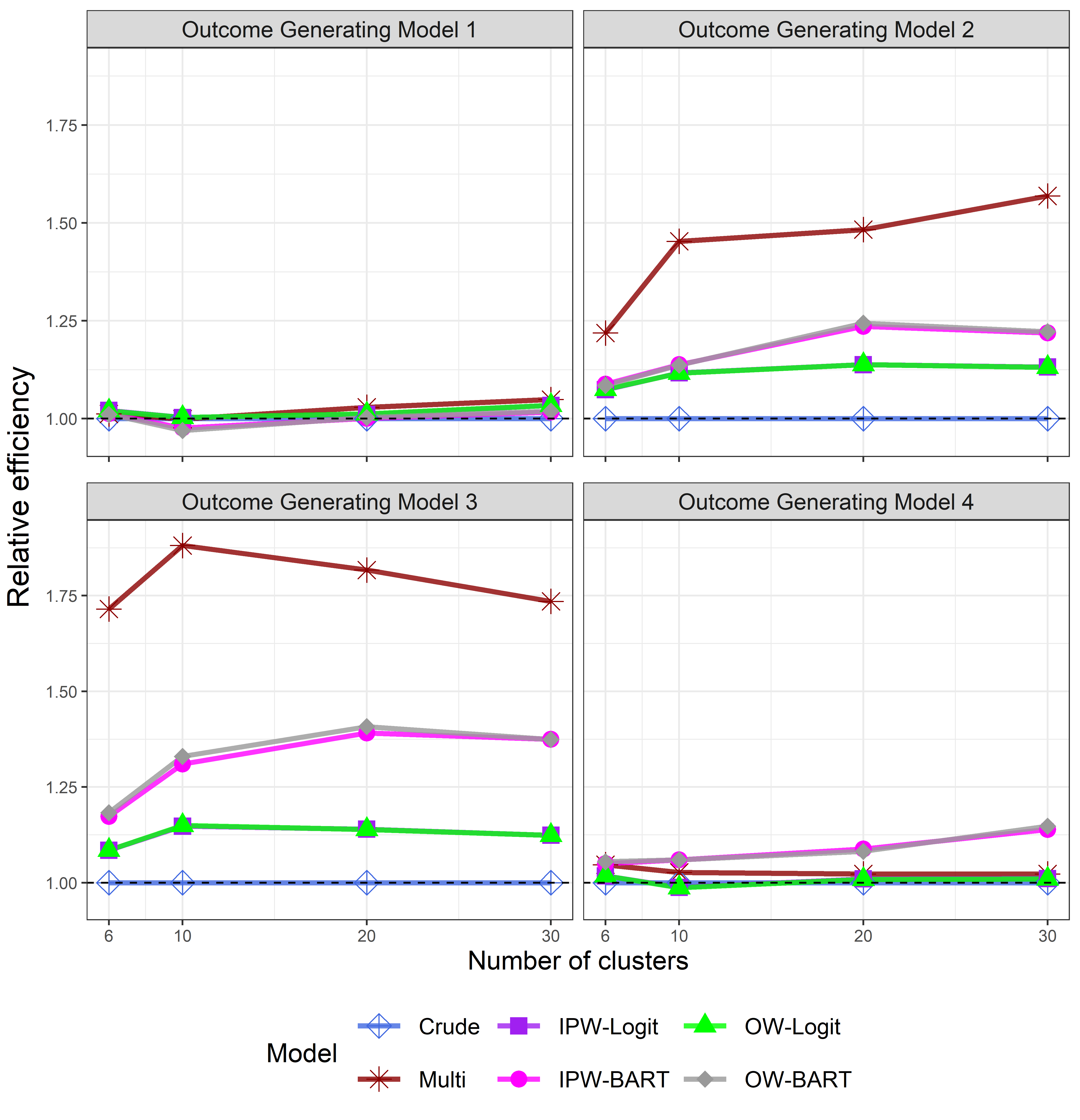}
    \caption{Measures of relative efficiency for 4 different simulation settings with average cluster size of 100 and ICC of 0.01 and assuming very low incidence. }
    \label{fig:RE}
\end{figure}

\subsection{Performance metrics under Outcome Generating Model 1}

\begin{table}[H]
\caption{Simulation results under Outcome Generating Model 1 with six covariates and low outcome incidences.}
\resizebox{\textwidth}{!}{
}   & Crude     & -0.844 & -0.012 & 1.000 & 0.931                                                  & 0.950   & 0.940    & 0                                                  \\
                                                             &                           & Multi     & -0.846 & -0.014 & 1.056 & 0.940                                                  & 0.951   & 0.947      & 0                                                  \\
                                                             &                           & IPW-Logit & -0.843 & -0.012 & 1.030 & 0.941                                                  & 0.954   & 0.948    & 0                                                  \\
                                                             &                           & IPW-BART  & -0.845 & -0.013 & 1.017 & 0.939                                                  & 0.957   & 0.950   & 0                                                  \\
                                                             &                           & OW-Logit  & -0.843 & -0.012 & 1.030 & 0.941                                                  & 0.955   & 0.948     & 0                                                  \\
                                                             &                           & OW-BART   & -0.844 & -0.013 & 1.016 & 0.942                                                  & 0.957   & 0.949    & 0                                                  \\ \hline
\end{tabular}
}
\end{table}

\begin{figure}[H]
    \centering
    \includegraphics[width=\linewidth]{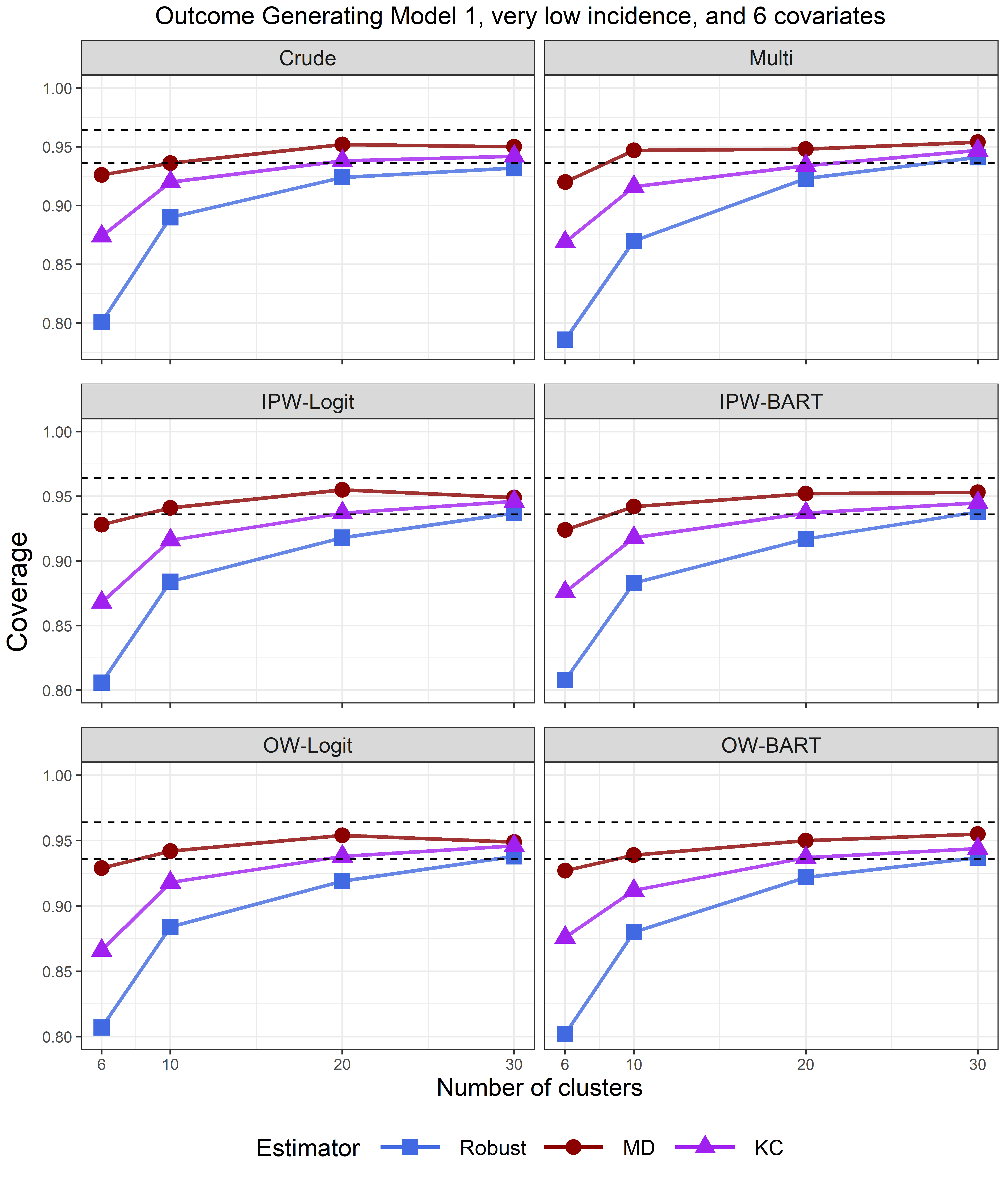}
    \caption{Measures of coverage for simulations based on the Outcome Generating Model 1, very low outcome incidences, 6 covariates, average cluster size of 100, and latent ICC of 0.01. }
    \label{fig:Out1VeryLowCov6m100ICC.01}
\end{figure}

\begin{table}[H]
\caption{Simulation results under Outcome Generating Model 1 with six covariates and very low outcome incidences. }
\resizebox{\textwidth}{!}{

}
\end{table}

\begin{table}[H]
\caption{Simulations results under Outcome Generating Model 1 with fifteen covariates and low outcome incidences. }
\resizebox{\textwidth}{!}{
%
}
\end{table}

\begin{table}[H]
\caption{Simulations results under Outcome Generating Model 1 with fifteen covariates and very low outcome incidences. }
\resizebox{\textwidth}{!}{
%
}   & Crude     & -0.831 & -0.016                   & 1.000 & 0.936                                                  & 0.950   & 0.945    & 0                                                  \\
                                                                                   &     & Multi     & -0.844 & -0.028                   & 1.057 & 0.933                                                  & 0.957   & 0.947    & 0                                                  \\
                                                                                    &    & IPW-Logit & -0.839 & -0.023                   & 1.029 & 0.942                                                  & 0.956   & 0.948    & 0                                                  \\
                                                                                    &    & IPW-BART  & -0.836 & -0.021                   & 1.020 & 0.937                                                  & 0.951   & 0.946     & 0                                                  \\
                                                                                    &    & OW-Logit  & -0.839 & -0.023                   & 1.037 & 0.941                                                  & 0.958   & 0.948    & 0                                                  \\
                                                                                    &    & OW-BART   & -0.838 & -0.022                   & 1.029 & 0.937                                                  & 0.953   & 0.948    & 0                                                  \\ \hline
\end{tabular}
}
\end{table}

\subsection{Performance metrics under Outcome Generating Model 2}

\begin{figure}[H]
    \centering
    \includegraphics[width=\linewidth]{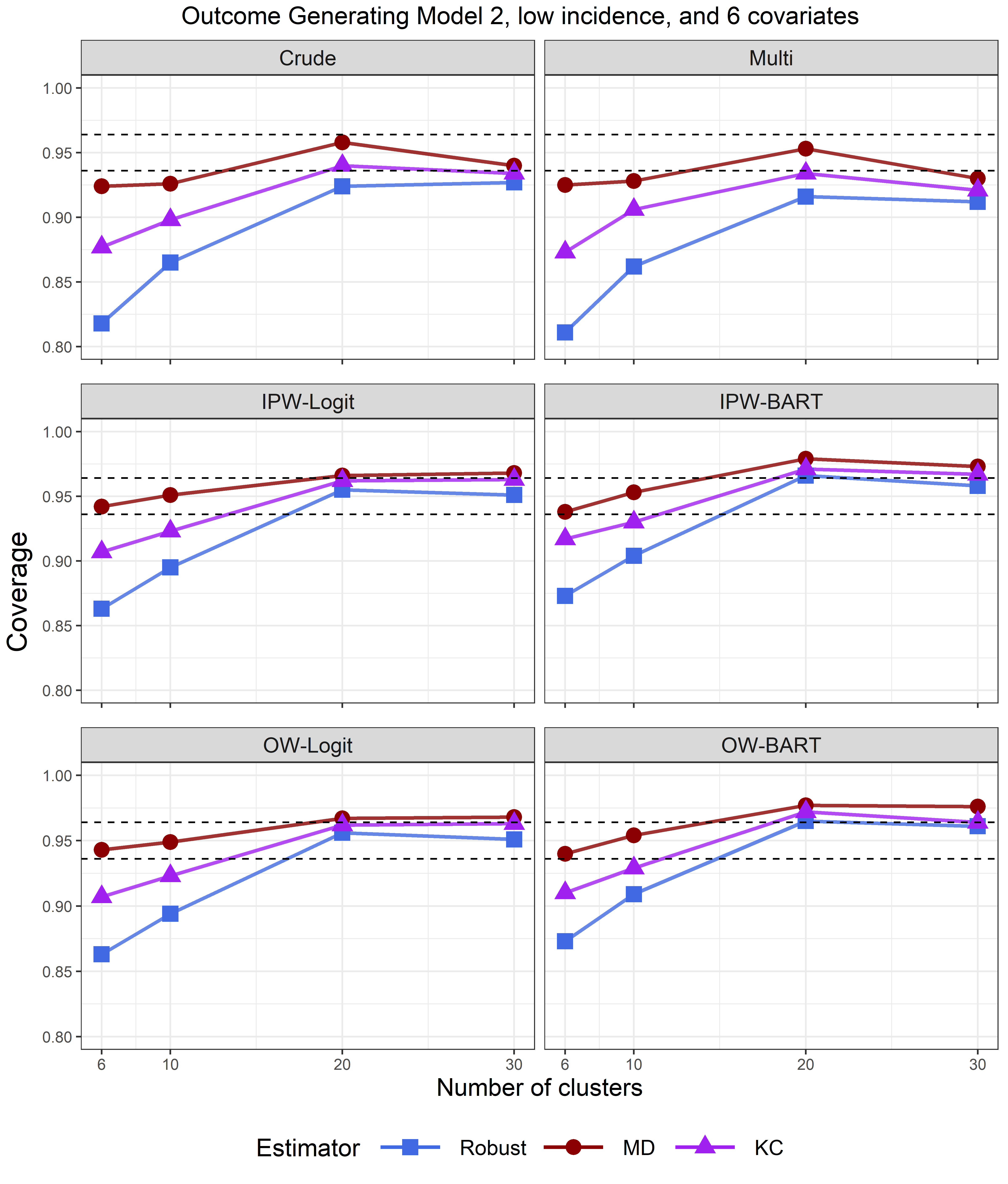}
    \caption{Measures of coverage for simulations based on Outcome Generating Model 2, low outcome incidences, 6 covariates, average cluster size of 100, and latent ICC of 0.01.}
    \label{fig:Out2LowCov6m100ICC.01}
\end{figure}

\begin{table}[H]
\caption{Simulations results under Outcome Generating Model 2 with six covariates and low outcome incidences. }
\resizebox{\textwidth}{!}{
}   & Crude     & -0.750 & -0.011 & 1.000  & 0.927                                                  & 0.948   & 0.938     & 0                                                  \\
                                                             &                           & Multi     & -0.755 & -0.016 & 1.683  & 0.930                                                  & 0.957   & 0.949     & 0                                                  \\
                                                             &                           & IPW-Logit & -0.751 & -0.011 & 1.268  & 0.945                                                  & 0.967   & 0.959  & 0                                                  \\
                                                             &                           & IPW-BART  & -0.753 & -0.014 & 1.319  & 0.956                                                  & 0.970   & 0.966     & 0                                                  \\
                                                             &                          & OW-Logit  & -0.751 & -0.012 & 1.268  & 0.946                                                  & 0.966   & 0.961    & 0                                                  \\
                                                             &                           & OW-BART   & -0.754 & -0.015 & 1.335  & 0.957                                                  & 0.971   & 0.966    & 0                                                  \\ \hline
\end{tabular}
}
\end{table}

\begin{figure}[H]
    \centering
    \includegraphics[width=\linewidth]{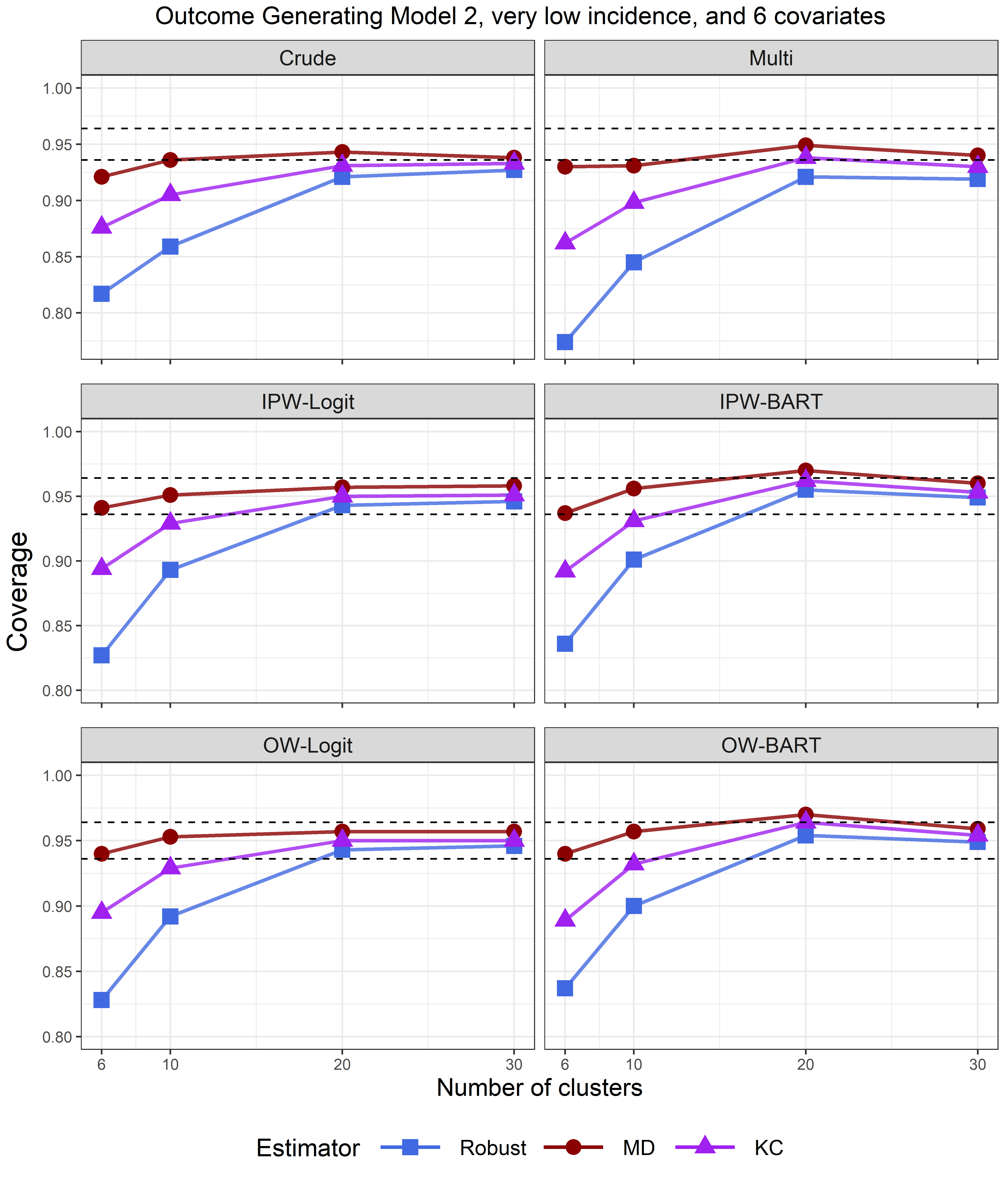}
    \caption{Measures of coverage for simulations based on Outcome Generating Model 2, very low outcome incidences, 6 covariates, average cluster size of 100, and latent ICC of 0.01. }
    \label{fig:Out2VeryLowCov6m100ICC.01}
\end{figure}

\begin{table}[H]
\caption{Simulations results under Outcome Generating Model 2 with six covariates and very low outcome incidences. }
\resizebox{\textwidth}{!}{

}
\end{table}

\begin{table}[H]
\caption{Simulations results under Outcome Generating Model 2 with fifteen covariates and low outcome incidences. }
\resizebox{\textwidth}{!}{
%
}
\end{table}

\begin{table}[H]
\caption{Simulations results under Outcome Generating Model 2 with fifteen covariates and very low outcome incidences. }
\resizebox{\textwidth}{!}{
%
}   & Crude     & -0.706 & -0.005 & 1.000 & 0.961                                                  & 0.977   & 0.970   & 0                                                  \\
                                                                                   &     & Multi     & -0.726 & -0.026 & 1.664 & 0.871                                                 & 0.978  & 0.935  & 0.003                                               \\
                                                                                   &     & IPW-Logit & -0.710 & -0.009 & 1.114 & 0.966                                                  & 0.976   & 0.969     & 0                                                  \\
                                                                                   &     & IPW-BART  & -0.710 & -0.010 & 1.103 & 0.967                                                  & 0.979   & 0.973   & 0                                                  \\
                                                                                   &     & OW-Logit  & -0.710 & -0.010 & 1.115 & 0.966                                                  & 0.975   & 0.970    & 0                                                  \\
                                                                                   &     & OW-BART   & -0.712 & -0.011 & 1.104 & 0.967                                                  & 0.979   & 0.973   & 0                                                  \\ \hline
\end{tabular}
}
\end{table}

\subsection{Performance metrics under Outcome Generating Model 3}

\begin{figure}[H]
    \centering
    \includegraphics[width=\linewidth]{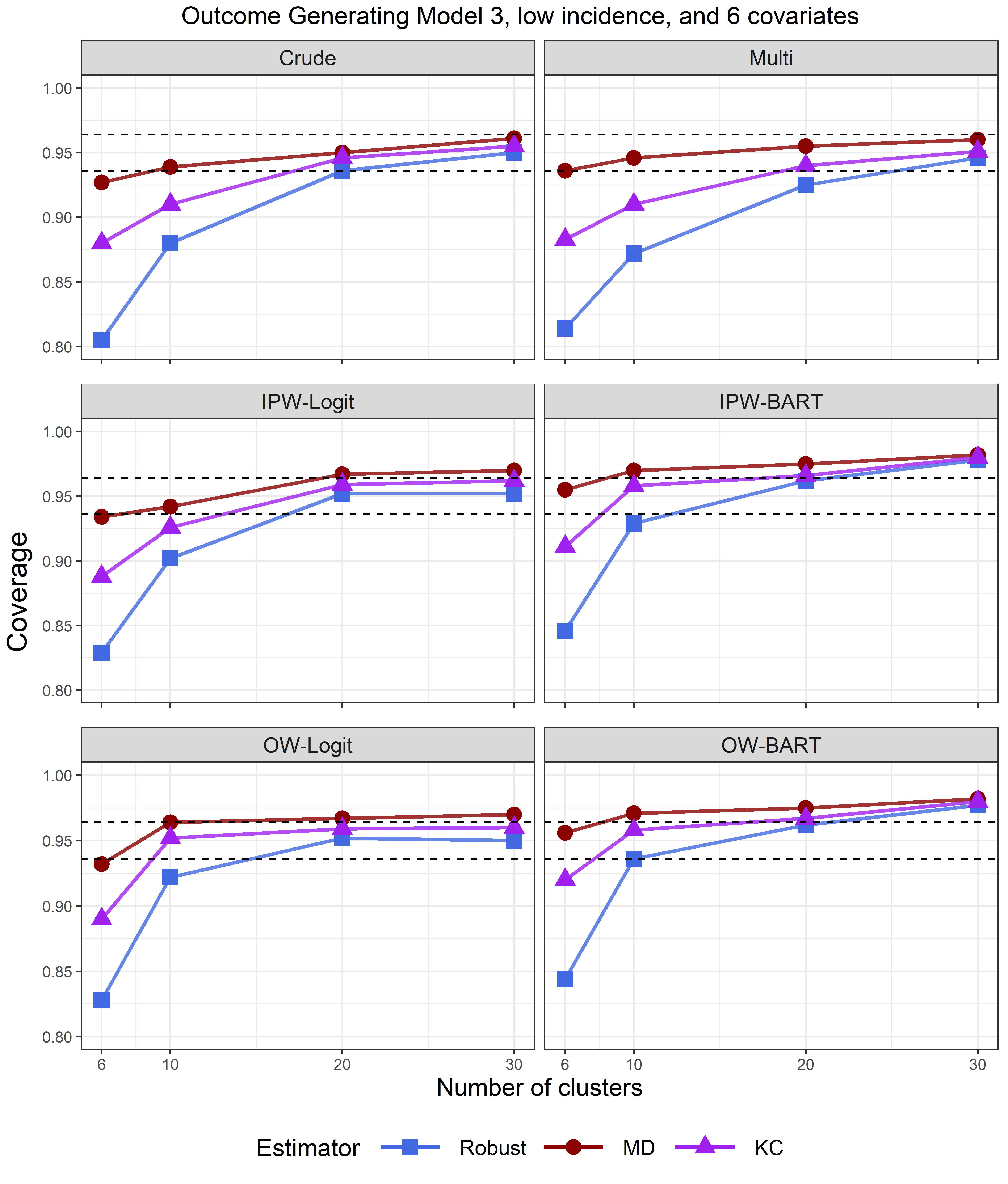}
    \caption{Measures of coverage for simulations based on Outcome Generating Model 3, low outcome incidences, average cluster size of 100 and latent ICC of 0.01. }
    \label{fig:Out3LowCov6m100ICC.01}
\end{figure}

\begin{table}[H]
\caption{Simulations results under the Outcome Generating Model 3 and low outcome incidences. }
\resizebox{\textwidth}{!}{
}   & Crude     & -0.765 & -0.026                   & 1.000 & 0.936                                                  & 0.954   & 0.943     & 0                                                  \\
                       &                                                                                         & Multi     & -0.762 & -0.023                   & 1.506 & 0.947                                                  & 0.963   & 0.957    & 0                                                  \\
                       &                                                                                         & IPW-Logit & -0.761 & -0.022                   & 1.129 & 0.948                                                  & 0.974   & 0.967    & 0                                                  \\
                       &                                                                                         & IPW-BART  & -0.762 & -0.023                   & 1.283 & 0.971                                                  & 0.979   & 0.974     & 0                                                  \\
                       &                                                                                         & OW-Logit  & -0.763 & -0.023                   & 1.125 & 0.957                                                  & 0.972   & 0.966    & 0                                                  \\
                       &                                                                                         & OW-BART   & -0.764 & -0.025                   & 1.292 & 0.971                                                  & 0.980   & 0.975  & 0                                                  \\ \hline
\end{tabular}
}
\end{table}

\begin{figure}[H]
    \centering
    \includegraphics[width=\linewidth]{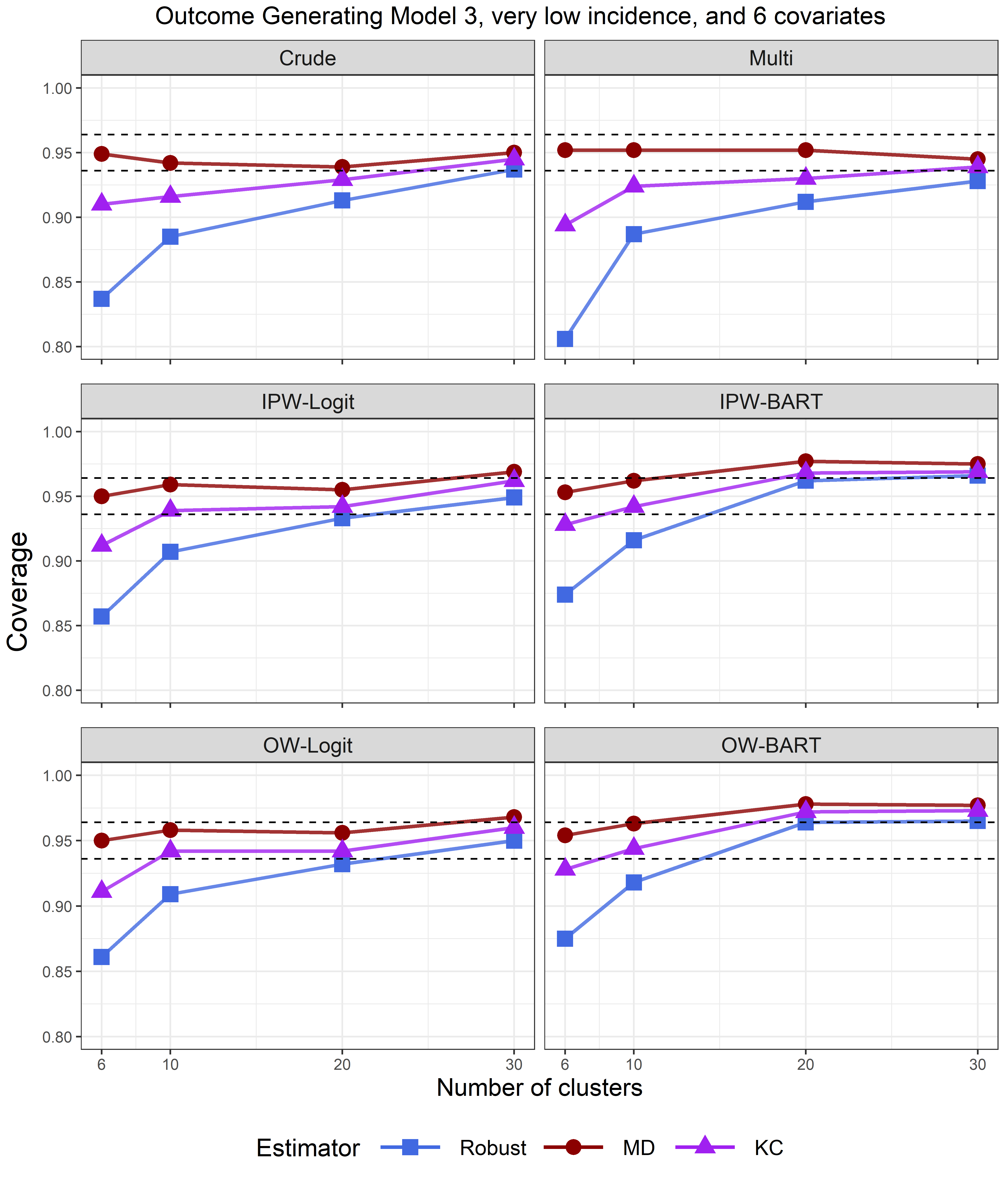}
    \caption{Measures of coverage for simulations based on Outcome Generating Model 3, very low outcome incidences, average cluster size of 100, and latent ICC of 0.01. }
    \label{fig:Out3VeryLowCov6m100ICC.01}
\end{figure}

\begin{table}[H]
\caption{Simulations results under the Outcome Generating Model 3 and very low outcome incidences. }
\resizebox{\textwidth}{!}{
}   & Crude     & -0.763 & -0.033 & 1.000 & 0.936                                                  & 0.952   & 0.943     & 0                                                  \\
                       &                                                                                         & Multi     & -0.770 & -0.040 & 1.790 & 0.929                                                  & 0.956   & 0.941     & 0                                                  \\
                       &                                                                                         & IPW-Logit & -0.760 & -0.030 & 1.129 & 0.954                                                  & 0.976   & 0.966    & 0                                                  \\
                       &                                                                                         & IPW-BART  & -0.765 & -0.035 & 1.293 & 0.966                                                  & 0.977   & 0.974    & 0                                                  \\
                       &                                                                                         & OW-Logit  & -0.762 & -0.032 & 1.127 & 0.953                                                  & 0.975   & 0.966   & 0                                                  \\
                       &                                                                                         & OW-BART   & -0.768 & -0.038 & 1.306 & 0.970                                                  & 0.980   & 0.975   & 0                                                  \\ \hline
\end{tabular}
}
\end{table}

\subsection{Performance metrics under Outcome Generating Model 4}

\begin{figure}[H]
    \centering
    \includegraphics[width=\linewidth]{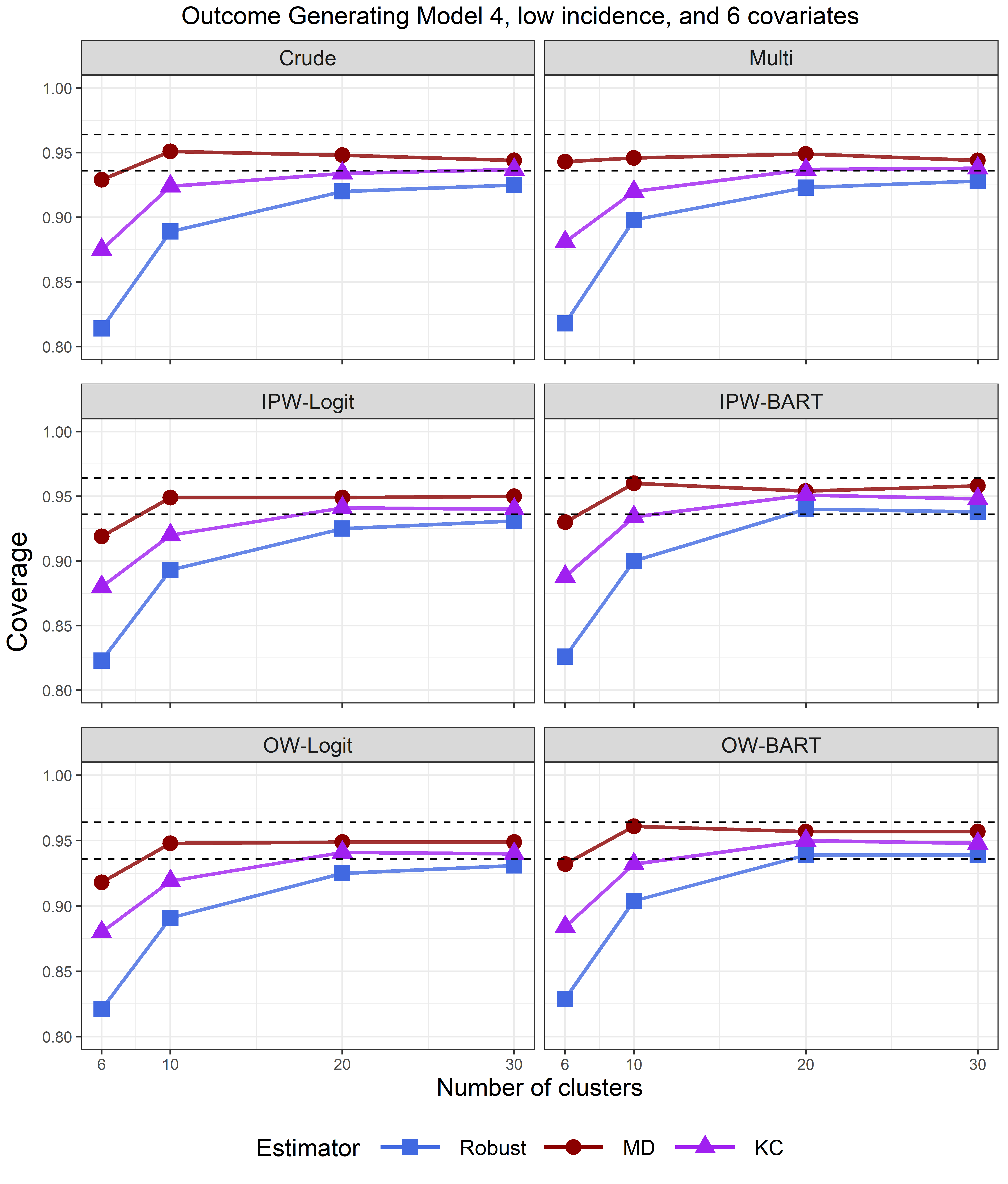}
    \caption{Measures of coverage for simulations based on Outcome Generating Model 4, low outcome incidences, average cluster size of 100, and latent ICC of 0.01. }
    \label{fig:Out4LowCov6m100ICC.01}
\end{figure}

\begin{table}[H]
\caption{Simulations results under the Outcome Generating Model 4 with six covariates and low outcome incidences. }
\resizebox{\textwidth}{!}{
}   & Crude     & -0.753 & -0.010                   & 1.000 & 0.942                                                  & 0.964   & 0.952    & 0                                                  \\
                       &                                                                                         & Multi     & -0.754 & -0.011                   & 1.012 & 0.935                                                  & 0.962   & 0.949   & 0                                                  \\
                       &                                                                                         & IPW-Logit & -0.753 & -0.010                   & 1.011 & 0.945                                                  & 0.962   & 0.954   &  0                                                  \\
                       &                                                                                         & IPW-BART  & -0.753 & -0.010                   & 1.068 & 0.945                                                  & 0.967   & 0.957   &  0                                                  \\
                       &                                                                                         & OW-Logit  & -0.754 & -0.011                   & 1.011 & 0.943                                                  & 0.961   & 0.955   &  0                                                  \\
                       &                                                                                         & OW-BART   & -0.755 & -0.012                   & 1.064 & 0.945                                                  & 0.967   & 0.957   &  0                                                  \\ \hline
\end{tabular}
}
\end{table}

\begin{figure}[H]
    \centering
    \includegraphics[width=\linewidth]{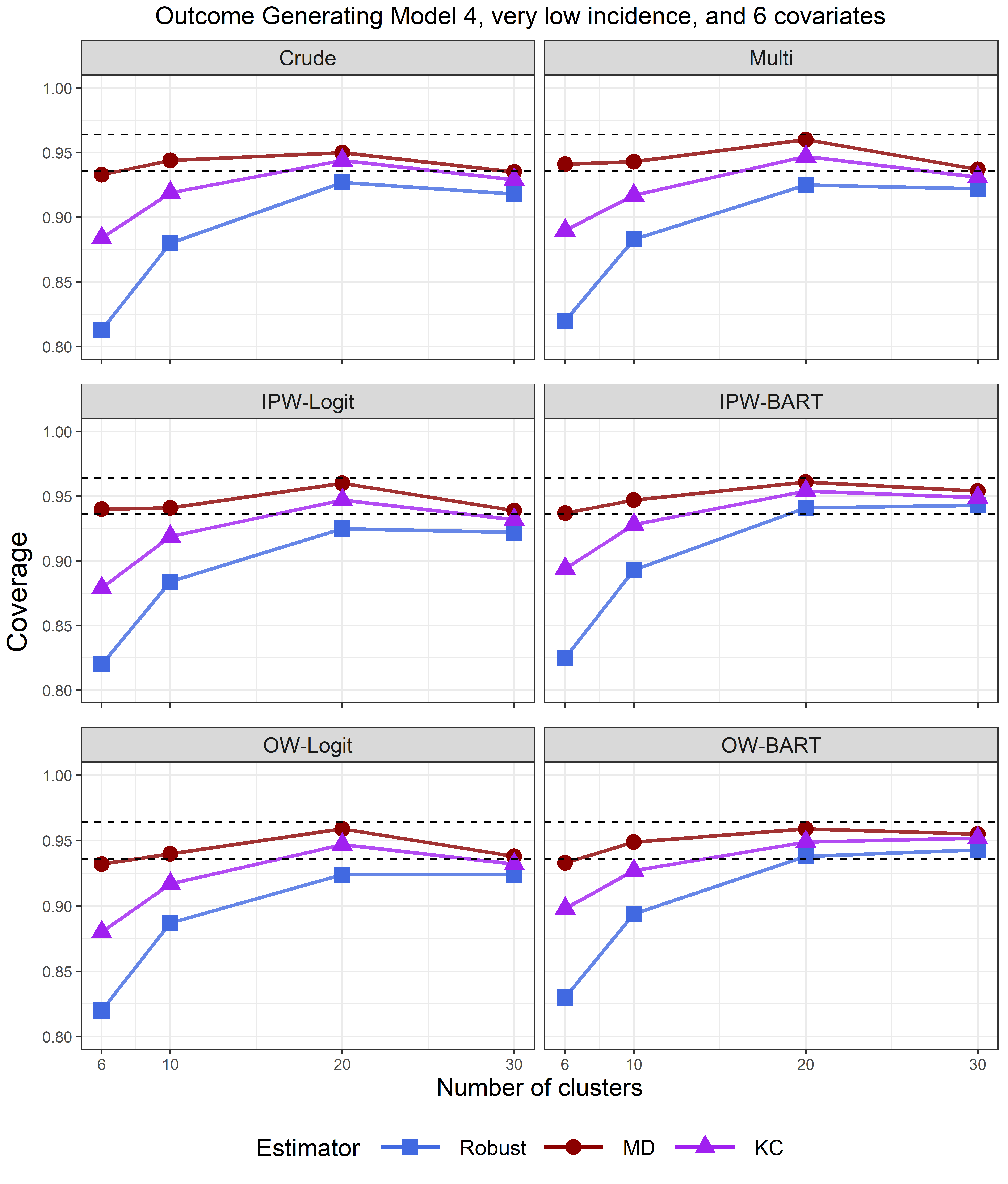}
    \caption{Measures of coverage for simulations based on Outcome Generating Model 4, very low outcome incidences, average cluster size of 100, and latent ICC of 0.01. }
    \label{fig:Out4VeryLowCov6m100ICC.01}
\end{figure}

\begin{table}[H]
\caption{Simulations results under the Outcome Generating Model 4 and very low outcome incidences. }
\resizebox{\textwidth}{!}{
}   & Crude     & -0.755 & -0.041                 & 1.000 & 0.938                                                  & 0.955   & 0.948   &  0                                                  \\
                       &                                                                                         & Multi     & -0.760 & -0.045                 & 1.025 & 0.933                                                  & 0.955   & 0.947   &  0                                                  \\
                       &                                                                                         & IPW-Logit & -0.755 & -0.041                 & 0.998  & 0.939                                                  & 0.955   & 0.947   & 0                                                  \\
                       &                                                                                         & IPW-BART  & -0.758 & -0.043                 & 1.035 & 0.951                                                  & 0.958   & 0.955   & 0                                                  \\
                       &                                                                                         & OW-Logit  & -0.756 & -0.042                 & 0.998  & 0.940                                                  & 0.955   & 0.948   &  0                                                  \\
                       &                                                                                         & OW-BART   & -0.759 & -0.045                 & 1.035 & 0.950                                                  & 0.961   & 0.955   &  0                                                  \\ \hline
\end{tabular}
}
\end{table}

\section{Data application}
\subsection{Assessing covariate balance}
\vspace{-4ex}
\begin{table}[H]
\caption{Covariate distributions obtained after propensity score weighting. ASD is the absolute standardized difference, a measure of covariate balance between the intervention and control groups. Cells contain mean (SD) for continuous variables and count (\%) for binary variables.}
\tiny
\resizebox{\textwidth}{!}{
\begin{tabular}{cccccc|ccccc}
\hline
\rowcolor[HTML]{DAE8FC} 
 &  &  & Control & Intervention & ASD &  &  & Control & Intervention & ASD \\ \hline
\cellcolor[HTML]{DAE8FC} &  & n & 2442.16 & 2459.09 &  &  & n & 2419.23 & 2419.13 &  \\
\cellcolor[HTML]{DAE8FC} &  & \cellcolor[HTML]{DAE8FC}Age & \cellcolor[HTML]{DAE8FC}4.79 (5.29) & \cellcolor[HTML]{DAE8FC}4.82 (5.69) & \cellcolor[HTML]{DAE8FC}0.005 &  & \cellcolor[HTML]{DAE8FC}Age & \cellcolor[HTML]{DAE8FC}4.71 (5.44) & \cellcolor[HTML]{DAE8FC}4.69 (5.45) & \cellcolor[HTML]{DAE8FC}0.003 \\
\cellcolor[HTML]{DAE8FC} &  & PRISM score & 9.04 (7.09) & 9.15 (8.25) & 0.014 &  & PRISM score & 8.95 (7.44) & 8.87 (7.47) & 0.010 \\
\cellcolor[HTML]{DAE8FC} &  & \cellcolor[HTML]{DAE8FC}\begin{tabular}[c]{@{}c@{}}Baseline POPC \\ score $= 1$\end{tabular} & \cellcolor[HTML]{DAE8FC}\begin{tabular}[c]{@{}c@{}}1732.2\\ (70.9)\end{tabular} & \cellcolor[HTML]{DAE8FC}\begin{tabular}[c]{@{}c@{}}1743.4\\ (70.9)\end{tabular} & \cellcolor[HTML]{DAE8FC}0.001 &  & \cellcolor[HTML]{DAE8FC}\begin{tabular}[c]{@{}c@{}}Baseline POPC\\ score $=1$\end{tabular} & \cellcolor[HTML]{DAE8FC}\begin{tabular}[c]{@{}c@{}}1727.4\\ (71.4)\end{tabular} & \cellcolor[HTML]{DAE8FC}\begin{tabular}[c]{@{}c@{}}1726.3\\ (71.4)\end{tabular} & \cellcolor[HTML]{DAE8FC}0.001 \\
\cellcolor[HTML]{DAE8FC} &  & Pneumonia & 870.6 (35.6) & 797.0 (32.4) & 0.068 &  & Pneumonia & 843.4 (34.9) & 814.1 (33.7) & 0.025 \\
\cellcolor[HTML]{DAE8FC} &  & \cellcolor[HTML]{DAE8FC}Bronchiolitis & \cellcolor[HTML]{DAE8FC}502.4 (20.6) & \cellcolor[HTML]{DAE8FC}775.6 (31.5) & \cellcolor[HTML]{DAE8FC}0.252 &  & \cellcolor[HTML]{DAE8FC}Bronchiolitis & \cellcolor[HTML]{DAE8FC}558.5 (23.1) & \cellcolor[HTML]{DAE8FC}724.9 (29.9) & \cellcolor[HTML]{DAE8FC}0.155 \\
\cellcolor[HTML]{DAE8FC} &  & \begin{tabular}[c]{@{}c@{}}Acute respiratory\\ failure\end{tabular} & 391.3 (16.0) & 340.4 (13.8) & 0.061 &  & \begin{tabular}[c]{@{}c@{}}Acute respiratory\\ failure\end{tabular} & 383.0 (15.8) & 328.1 (13.6) & 0.064 \\
\cellcolor[HTML]{DAE8FC} &  & \cellcolor[HTML]{DAE8FC}\begin{tabular}[c]{@{}c@{}}Oxygenation\\ index\end{tabular} & \cellcolor[HTML]{DAE8FC}8.11 (7.12) & \cellcolor[HTML]{DAE8FC}8.40 (7.08) & \cellcolor[HTML]{DAE8FC}0.041 &  & \cellcolor[HTML]{DAE8FC}\begin{tabular}[c]{@{}c@{}}Oxygenation\\ index\end{tabular} & \cellcolor[HTML]{DAE8FC}8.03 (7.02) & \cellcolor[HTML]{DAE8FC}8.40 (7.02) & \cellcolor[HTML]{DAE8FC}0.052 \\
\cellcolor[HTML]{DAE8FC} &  & Prematurity & 361.3 (13.8) & 370.0 (15.0) & 0.007 &  & Prematurity & 351.0 (14.5) & 367.5 (15.2) & 0.019 \\
\cellcolor[HTML]{DAE8FC} &  & \cellcolor[HTML]{DAE8FC}Asthma & \cellcolor[HTML]{DAE8FC}413.7 (16.9) & \cellcolor[HTML]{DAE8FC}299.9 (12.2) & \cellcolor[HTML]{DAE8FC}0.135 &  & \cellcolor[HTML]{DAE8FC}Asthma & \cellcolor[HTML]{DAE8FC}380.2 (15.7) & \cellcolor[HTML]{DAE8FC}319.9 (13.2) & \cellcolor[HTML]{DAE8FC}0.071 \\
\cellcolor[HTML]{DAE8FC} &  & Cancer & 196.5 (8.0) & 205.2 (8.3) & 0.011 &  & Cancer & 189.1 (7.8) & 207.7 (8.6) & 0.028 \\
\cellcolor[HTML]{DAE8FC} & \multirow{-12}{*}{\begin{tabular}[c]{@{}c@{}}IPW-\\ Logit\end{tabular}} & \cellcolor[HTML]{DAE8FC}Transferred & \cellcolor[HTML]{DAE8FC}592.1 (24.2) & \cellcolor[HTML]{DAE8FC}681.7 (27.7) & \cellcolor[HTML]{DAE8FC}0.079 & \multirow{-12}{*}{\begin{tabular}[c]{@{}c@{}}IPW-\\ BART\end{tabular}} & \cellcolor[HTML]{DAE8FC}Transferred & \cellcolor[HTML]{DAE8FC}581.8 (24.1) & \cellcolor[HTML]{DAE8FC}673.3 (27.8) & \cellcolor[HTML]{DAE8FC}0.086 \\ \cline{2-10} 
\cellcolor[HTML]{DAE8FC} &  & n & 599.65 & 599.65 &  &  & n & 575.30 & 576.10 &  \\
\cellcolor[HTML]{DAE8FC} &  & \cellcolor[HTML]{DAE8FC}Age & \cellcolor[HTML]{DAE8FC}4.74 (5.25) & \cellcolor[HTML]{DAE8FC}4.74 (5.65) & \cellcolor[HTML]{DAE8FC}$<0.001$ &  & \cellcolor[HTML]{DAE8FC}Age & \cellcolor[HTML]{DAE8FC}4.76 (5.45) & \cellcolor[HTML]{DAE8FC}4.76 (5.46) & \cellcolor[HTML]{DAE8FC}$<0.001$ \\
\cellcolor[HTML]{DAE8FC} &  & PRISM score & 8.92 (6.91) & 8.92 (7.89) & $<0.001$ &  & PRISM score & 9.02 (7.42) & 8.99 (7.46) & 0.004 \\
\cellcolor[HTML]{DAE8FC} &  & \cellcolor[HTML]{DAE8FC}\begin{tabular}[c]{@{}c@{}}Baseline POPC \\ score $=1$\end{tabular} & \cellcolor[HTML]{DAE8FC}\begin{tabular}[c]{@{}c@{}}425.4\\ (70.9)\end{tabular} & \cellcolor[HTML]{DAE8FC}\begin{tabular}[c]{@{}c@{}}425.4\\ (70.9)\end{tabular} & \cellcolor[HTML]{DAE8FC}$<0.001$ &  & \cellcolor[HTML]{DAE8FC}\begin{tabular}[c]{@{}c@{}}Baseline POPC \\ score $= 1$\end{tabular} & \cellcolor[HTML]{DAE8FC}\begin{tabular}[c]{@{}c@{}}406.5\\ (70.7)\end{tabular} & \cellcolor[HTML]{DAE8FC}\begin{tabular}[c]{@{}c@{}}406.8\\ (70.6)\end{tabular} & \cellcolor[HTML]{DAE8FC}0.001 \\
\cellcolor[HTML]{DAE8FC} &  & Pneumonia & 214.3 (35.7) & 195.6 (32.6) & 0.066 &  & Pneumonia & 203.3 (35.3) & 195.8 (34.0) & 0.028 \\
\cellcolor[HTML]{DAE8FC} &  & \cellcolor[HTML]{DAE8FC}Bronchiolitis & \cellcolor[HTML]{DAE8FC}123.8 (20.6) & \cellcolor[HTML]{DAE8FC}190.6 (31.8) & \cellcolor[HTML]{DAE8FC}0.255 &  & \cellcolor[HTML]{DAE8FC}Bronchiolitis & \cellcolor[HTML]{DAE8FC}129.1 (22.4) & \cellcolor[HTML]{DAE8FC}168.0 (29.2) & \cellcolor[HTML]{DAE8FC}0.154 \\
\cellcolor[HTML]{DAE8FC} &  & \begin{tabular}[c]{@{}c@{}}Acute respiratory\\ failure\end{tabular} & 95.0 (15.8) & 81.0 (13.5) & 0.066 &  & \begin{tabular}[c]{@{}c@{}}Acute respiratory\\ failure\end{tabular} & 91.5 (15.9) & 79.0 (13.7) & 0.062 \\
\cellcolor[HTML]{DAE8FC} &  & \cellcolor[HTML]{DAE8FC}\begin{tabular}[c]{@{}c@{}}Oxygenation\\ index\end{tabular} & \cellcolor[HTML]{DAE8FC}8.09 (7.10) & \cellcolor[HTML]{DAE8FC}8.36 (7.05) & \cellcolor[HTML]{DAE8FC}0.038 &  & \cellcolor[HTML]{DAE8FC}\begin{tabular}[c]{@{}c@{}}Oxygenation\\ index\end{tabular} & \cellcolor[HTML]{DAE8FC}8.04 (7.02) & \cellcolor[HTML]{DAE8FC}8.44 (7.06) & \cellcolor[HTML]{DAE8FC}0.056 \\
\cellcolor[HTML]{DAE8FC} &  & Prematurity & 89.1 (14.9) & 91.0 (15.2) & 0.009 &  & Prematurity & 84.5 (14.7) & 87.9 (15.2) & 0.016 \\
\cellcolor[HTML]{DAE8FC} &  & \cellcolor[HTML]{DAE8FC}Asthma & \cellcolor[HTML]{DAE8FC}102.1 *17.0) & \cellcolor[HTML]{DAE8FC}73.7 (12.3) & \cellcolor[HTML]{DAE8FC}0.134 &  & \cellcolor[HTML]{DAE8FC}Asthma & \cellcolor[HTML]{DAE8FC}91.4 (15.9) & \cellcolor[HTML]{DAE8FC}77.1 (13.4) & \cellcolor[HTML]{DAE8FC}0.071 \\
\cellcolor[HTML]{DAE8FC} &  & Cancer & 47.7 (7.9) & 49.2 (8.2) & 0.010 &  & Cancer & 45.3 (7.9) & 49.7 (8.6) & 0.027 \\
\multirow{-24}{*}{\cellcolor[HTML]{DAE8FC}\begin{tabular}[c]{@{}c@{}}Covariate\\ adjust-\\ ment\\ set 1\end{tabular}} & \multirow{-12}{*}{\begin{tabular}[c]{@{}c@{}}OW-\\ Logit\end{tabular}} & \cellcolor[HTML]{DAE8FC}Transferred & \cellcolor[HTML]{DAE8FC}144.8 (24.2) & \cellcolor[HTML]{DAE8FC}166.2 (27.7) & \cellcolor[HTML]{DAE8FC}0.081 & \multirow{-12}{*}{\begin{tabular}[c]{@{}c@{}}OW-\\ BART\end{tabular}} & \cellcolor[HTML]{DAE8FC}Transferred & \cellcolor[HTML]{DAE8FC}139.5 (24.3) & \cellcolor[HTML]{DAE8FC}160.8 (27.9) & \cellcolor[HTML]{DAE8FC}0.083 \\ \hline
\cellcolor[HTML]{DAE8FC} &  & n & 2408.47 & 2415.19 &  &  & n & 2377.83 & 2369.52 &  \\
\cellcolor[HTML]{DAE8FC} &  & \cellcolor[HTML]{DAE8FC}Age & \cellcolor[HTML]{DAE8FC}4.67 (5.33) & \cellcolor[HTML]{DAE8FC}4.68 (5.53) & \cellcolor[HTML]{DAE8FC}0.001 &  & \cellcolor[HTML]{DAE8FC}Age & \cellcolor[HTML]{DAE8FC}4.63 (5.40) & \cellcolor[HTML]{DAE8FC}4.61 (5.41) & \cellcolor[HTML]{DAE8FC}0.003 \\
\cellcolor[HTML]{DAE8FC} &  & PRISM score & 8.88 (7.04) & 8.94 (7.98) & 0.008 &  & PRISM score & 8.82 (7.32) & 8.73 (7.39) & 0.012 \\
\cellcolor[HTML]{DAE8FC} &  & \cellcolor[HTML]{DAE8FC}\begin{tabular}[c]{@{}c@{}}Baseline POPC \\ score $= 1$\end{tabular} & \cellcolor[HTML]{DAE8FC}\begin{tabular}[c]{@{}c@{}}1702.0\\ (70.7)\end{tabular} & \cellcolor[HTML]{DAE8FC}\begin{tabular}[c]{@{}c@{}}1712.4\\ (70.9)\end{tabular} & \cellcolor[HTML]{DAE8FC}0.005 &  & \cellcolor[HTML]{DAE8FC}\begin{tabular}[c]{@{}c@{}}Baseline POPC \\ score $= 1$\end{tabular} & \cellcolor[HTML]{DAE8FC}\begin{tabular}[c]{@{}c@{}}1692.2\\ (71.2)\end{tabular} & \cellcolor[HTML]{DAE8FC}\begin{tabular}[c]{@{}c@{}}1687.3\\ (71.2)\end{tabular} & \cellcolor[HTML]{DAE8FC}0.001 \\
\cellcolor[HTML]{DAE8FC} &  & Pneumonia & 812.7 (33.7) & 812.9 (33.7) & 0.002 &  & Pneumonia & 805.7 (33.9) & 806.1 (34.0) & 0.003 \\
\cellcolor[HTML]{DAE8FC} &  & \cellcolor[HTML]{DAE8FC}Bronchiolitis & \cellcolor[HTML]{DAE8FC}652.8 (27.1) & \cellcolor[HTML]{DAE8FC}654.0 (27.1) & \cellcolor[HTML]{DAE8FC}0.001 &  & \cellcolor[HTML]{DAE8FC}Bronchiolitis & \cellcolor[HTML]{DAE8FC}645.2 (27.1) & \cellcolor[HTML]{DAE8FC}652.3 (27.5) & \cellcolor[HTML]{DAE8FC}0.009 \\
\cellcolor[HTML]{DAE8FC} &  & \begin{tabular}[c]{@{}c@{}}Acute respiratory\\ failure\end{tabular} & 347.6 (14.4) & 350.5 (14.5) & 0.002 &  & \begin{tabular}[c]{@{}c@{}}Acute respiratory\\ failure\end{tabular} & 339.4 (14.3) & 331.6 (14.0) & 0.008 \\
\cellcolor[HTML]{DAE8FC} &  & \cellcolor[HTML]{DAE8FC}\begin{tabular}[c]{@{}c@{}}Oxygenation\\ index\end{tabular} & \cellcolor[HTML]{DAE8FC}8.25 (7.34) & \cellcolor[HTML]{DAE8FC}8.24 (6.82) & \cellcolor[HTML]{DAE8FC}0.003 &  & \cellcolor[HTML]{DAE8FC}\begin{tabular}[c]{@{}c@{}}Oxygenation\\ index\end{tabular} & \cellcolor[HTML]{DAE8FC}8.16 (7.02) & \cellcolor[HTML]{DAE8FC}8.16 (6.83) & \cellcolor[HTML]{DAE8FC}0.001 \\
\cellcolor[HTML]{DAE8FC} &  & Prematurity & 363.0 (15.1) & 369.1 (15.3) & 0.006 &  & Prematurity & 351.5 (14.8) & 361.1 (15.2) & 0.013 \\
\cellcolor[HTML]{DAE8FC} &  & \cellcolor[HTML]{DAE8FC}Asthma & \cellcolor[HTML]{DAE8FC}391.5 (16.3) & \cellcolor[HTML]{DAE8FC}309.3 (12.8) & \cellcolor[HTML]{DAE8FC}0.098 &  & \cellcolor[HTML]{DAE8FC}Asthma & \cellcolor[HTML]{DAE8FC}370.8 (15.6) & \cellcolor[HTML]{DAE8FC}320.8 (13.5) & \cellcolor[HTML]{DAE8FC}0.058 \\
\cellcolor[HTML]{DAE8FC} &  & Cancer & 191.9 (8.0) & 205.2 (8.5) & 0.019 &  & Cancer & 186.2 (7.8) & 195.4 (8.2) & 0.015 \\
\cellcolor[HTML]{DAE8FC} & \multirow{-12}{*}{\begin{tabular}[c]{@{}c@{}}IPW-\\ Logit\end{tabular}} & \cellcolor[HTML]{DAE8FC}Transferred & \cellcolor[HTML]{DAE8FC}579.9 (24.1) & \cellcolor[HTML]{DAE8FC}662.2 (27.4) & \cellcolor[HTML]{DAE8FC}0.077 & \multirow{-12}{*}{\begin{tabular}[c]{@{}c@{}}IPW-\\ BART\end{tabular}} & \cellcolor[HTML]{DAE8FC}Transferred & \cellcolor[HTML]{DAE8FC}566.6 (23.8) & \cellcolor[HTML]{DAE8FC}656.5 (27.7) & \cellcolor[HTML]{DAE8FC}0.089 \\ \cline{2-10} 
\cellcolor[HTML]{DAE8FC} &  & n & 576.59 & 576.59 &  &  & n & 559.48 & 558.87 &  \\
\cellcolor[HTML]{DAE8FC} &  & \cellcolor[HTML]{DAE8FC}Age & \cellcolor[HTML]{DAE8FC}4.73 (5.34) & \cellcolor[HTML]{DAE8FC}4.73 (5.54) & \cellcolor[HTML]{DAE8FC}$<0.001$ &  & \cellcolor[HTML]{DAE8FC}Age & \cellcolor[HTML]{DAE8FC}4.70 (5.42) & \cellcolor[HTML]{DAE8FC}4.70 (5.43) & \cellcolor[HTML]{DAE8FC}0 \\
\cellcolor[HTML]{DAE8FC} &  & PRISM score & 8.84 (6.92) & 8.84 (7.74) & $<0.001$ &  & PRISM score & 8.90 (7.29) & 8.85 (7.38) & 0.006 \\
\cellcolor[HTML]{DAE8FC} &  & \cellcolor[HTML]{DAE8FC}\begin{tabular}[c]{@{}c@{}}Baseline POPC \\ score $=1$\end{tabular} & \cellcolor[HTML]{DAE8FC}\begin{tabular}[c]{@{}c@{}}406.1\\ (70.4)\end{tabular} & \cellcolor[HTML]{DAE8FC}\begin{tabular}[c]{@{}c@{}}406.1\\ (70.4)\end{tabular} & \cellcolor[HTML]{DAE8FC}$<0.001$ &  & \cellcolor[HTML]{DAE8FC}\begin{tabular}[c]{@{}c@{}}Baseline POPC \\ score $=1$\end{tabular} & \cellcolor[HTML]{DAE8FC}\begin{tabular}[c]{@{}c@{}}393.5\\ (70.3)\end{tabular} & \cellcolor[HTML]{DAE8FC}\begin{tabular}[c]{@{}c@{}}393.0\\ (70.3)\end{tabular} & \cellcolor[HTML]{DAE8FC}$<0.001$ \\
\cellcolor[HTML]{DAE8FC} &  & Pneumonia & 201.4 (34.9) & 201.4 (34.9) & \cellcolor[HTML]{FFFFFF}$<0.001$ &  & Pneumonia & 197.0 (35.2) & 196.7 (35.2) & $<0.001$ \\
\cellcolor[HTML]{DAE8FC} &  & \cellcolor[HTML]{DAE8FC}Bronchiolitis & \cellcolor[HTML]{DAE8FC}147.7 (25.6) & \cellcolor[HTML]{DAE8FC}147.7 (25.6) & \cellcolor[HTML]{DAE8FC}$<0.001$ &  & \cellcolor[HTML]{DAE8FC}Bronchiolitis & \cellcolor[HTML]{DAE8FC}144.0 (25.7) & \cellcolor[HTML]{DAE8FC}144.3 (25.8) & \cellcolor[HTML]{DAE8FC}0.002 \\
\cellcolor[HTML]{DAE8FC} &  & \begin{tabular}[c]{@{}c@{}}Acute respiratory\\ failure\end{tabular} & 83.1 (14.4) & 83.1 (14.4) & \cellcolor[HTML]{FFFFFF}$<0.001$ &  & \begin{tabular}[c]{@{}c@{}}Acute respiratory\\ failure\end{tabular} & 79.3 (14.2) & 78.9 (14.1) & 0.002 \\
\cellcolor[HTML]{DAE8FC} &  & \cellcolor[HTML]{DAE8FC}\begin{tabular}[c]{@{}c@{}}Oxygenation\\ index\end{tabular} & \cellcolor[HTML]{DAE8FC}8.26 (7.35) & \cellcolor[HTML]{DAE8FC}8.26 (6.86) & \cellcolor[HTML]{DAE8FC}$<0.001$ &  & \cellcolor[HTML]{DAE8FC}\begin{tabular}[c]{@{}c@{}}Oxygenation\\ index\end{tabular} & \cellcolor[HTML]{DAE8FC}8.20 (7.00) & \cellcolor[HTML]{DAE8FC}8.23 (6.88) & \cellcolor[HTML]{DAE8FC}0.004 \\
\cellcolor[HTML]{DAE8FC} &  & Prematurity & 86.5 (15.0) & 87.9 (15.3) & 0.007 &  & Prematurity & 83.4 (14.9) & 85.0 (15.2) & 0.008 \\
\cellcolor[HTML]{DAE8FC} &  & \cellcolor[HTML]{DAE8FC}Asthma & \cellcolor[HTML]{DAE8FC}94.6 (16.4) & \cellcolor[HTML]{DAE8FC}74.9 (13.0) & \cellcolor[HTML]{DAE8FC}0.096 &  & \cellcolor[HTML]{DAE8FC}Asthma & \cellcolor[HTML]{DAE8FC}87.8 (15.7) & \cellcolor[HTML]{DAE8FC}75.9 (13.6) & \cellcolor[HTML]{DAE8FC}0.060 \\
\cellcolor[HTML]{DAE8FC} &  & Cancer & 46.2 (8.0) & 49.0 (8.5) & 0.018 &  & Cancer & 44.5 (8.0) & 46.9 (8.4) & 0.016 \\
\multirow{-24}{*}{\cellcolor[HTML]{DAE8FC}\begin{tabular}[c]{@{}c@{}}Covariate\\ adjust-\\ ment\\ set 2\end{tabular}} & \multirow{-12}{*}{\begin{tabular}[c]{@{}c@{}}OW-\\ Logit\end{tabular}} & \cellcolor[HTML]{DAE8FC}Transferred & \cellcolor[HTML]{DAE8FC}148.2 (25.8) & \cellcolor[HTML]{DAE8FC}148.2 (25.8) & \cellcolor[HTML]{DAE8FC}$<0.001$ & \multirow{-12}{*}{\begin{tabular}[c]{@{}c@{}}OW-\\ BART\end{tabular}} & \cellcolor[HTML]{DAE8FC}Transferred & \cellcolor[HTML]{DAE8FC}143.0 (25.6) & \cellcolor[HTML]{DAE8FC}143.1 (25.6) & \cellcolor[HTML]{DAE8FC}$<0.001$ \\ \hline
\end{tabular}
}
\end{table}

%\subsection{Effect estimates for other outcomes of interest}

%\begin{figure}[H]
%    \centering
%    \includegraphics[width=\linewidth]{RESTORE_OutcomeA.png}
%    \caption{Estimates of participant-average treatment effect and 95\% confidence intervals (CIs) for not successfully extubated outcome.}
%    \label{fig:restoreA}
%\end{figure}

%\begin{figure}[H]
%    \centering
%    \includegraphics[width=\linewidth]{RESTORE_OutcomeB.png}
%    \caption{Estimates of participant-average treatment effect and 95\% confidence intervals (CIs) for 90-day mortality outcome.}
%    \label{fig:restoreB}
%\end{figure}